\documentclass[twoside,11pt]{article}

\usepackage{jmlr2e_pre}

\jmlrheading{}{2022}{1-24}{8/22}{}{}{Yoshiyuki Ninomiya}

\ShortHeadings{Prior Intensified Information Criterion}{Y. Ninomiya}
\firstpageno{1}

\usepackage{amsmath,latexsym,amssymb}
\usepackage{graphicx,bm,fancybox,multirow,hhline,indentfirst}
\usepackage{ascmac,atbegshi}
\usepackage{natbib,verbatim,color}
\usepackage{booktabs,subfig,url}

\newcommand{\argmax}{\mathop{\rm argmax}}

\newcommand{\indep}{\mathop{\perp\!\!\!\!\perp}}

\selectfont

\def\T{{\rm T}}
\def\E{{\rm E}}
\def\P{{\rm P}}
\def\O{{\rm O}}

\def\oP{{\rm o}_{\rm P}}
\def\OP{{\rm O}_{\rm P}}

\begin{document}

\title{Prior Intensified Information Criterion}

\author{\name Yoshiyuki Ninomiya \email ninomiya@ism.ac.jp\\
\addr Department of Statistical Inference and Mathematics\\
The Institute of Statistical Mathematics\\
10-3 Midori-cho, Tachikawa-shi, Tokyo 190-8562, Japan}

\editor{}

\maketitle

\begin{abstract}
The widely applicable information criterion (WAIC) has been used as a model selection criterion for Bayesian statistics in recent years. It is an asymptotically unbiased estimator of the Kullback-Leibler divergence between a Bayesian predictive distribution and the true distribution. Not only is the WAIC theoretically more sound than other information criteria, its usefulness in practice has also been reported. On the other hand, the WAIC is intended for settings in which the prior distribution does not have an asymptotic influence, and as we set the class of the prior distribution to be more complex, it never fails to select the most complex one. To alleviate these concerns, this paper proposed the prior intensified information criterion (PIIC). In addition, it customizes this criterion to incorporate sparse estimation and causal inference. Numerical experiments show that the PIIC clearly outperforms the WAIC in terms of prediction performance when the above concerns are manifested. A real data analysis confirms that the results of variable selection and Bayesian estimators of the WAIC and PIIC differ significantly.
\end{abstract}


\section{Introduction}
\label{sec1}
The deviance information criterion (DIC) proposed by \cite{SpiBCL02} was a breakthrough in Bayesian modeling. The DIC is a Bayesian criterion, but it assumes there is a true distribution and evaluates the Kullback-Leibler divergence from the true distribution, as in Akaike information criterion (AIC; \citealt{Aka73}). Let $\bm{z}=(z_1,z_2,\ldots,z_n)$ be samples from the probability function $f(\cdot\mid\bm{\theta})$ and $\bm{\xi}$ be a hyper-parameter of the prior distribution. Further, let us denote the posterior expectation of $\bm{\theta}$ by $\E_{\bm{\theta}\mid\bm{z};\bm{\xi}}(\cdot)$. Then, the DIC is written as
\begin{align}
\sum_{i=1}^n\log f\{z_i\mid\E_{\bm{\theta}\mid\bm{z};\bm{\xi}}(\bm{\theta})\}
-2\sum_{i=1}^n\E_{\bm{\theta}\mid\bm{z};\bm{\xi}}\{\log f(z_i\mid\bm{\theta})\}.
\label{dic}
\end{align}
Since it is a quantity that can be obtained when the posterior expectation is evaluated, it can be implemented using a Markov chain Monte Carlo (MCMC) method.

The widely applicable information criterion (WAIC) was derived by \cite{Wat10} as an asymptotically unbiased estimator of the Kullback-Leibler divergence in a general framework. In fact, in the settings addressed in that paper and in \cite{Wat15}, the WAIC clearly gave experimental results that were more valid than those of the DIC. Moreover, even though the WAIC was derived using difficult algebraic geometry tools, its final form is simple:
\begin{align}
-\sum_{i=1}^n\log\E_{\bm{\theta}\mid\bm{z};\bm{\xi}}\{f(z_i\mid\bm{\theta})\}
+\sum_{i=1}^n\E_{\bm{\theta}\mid\bm{z};\bm{\xi}}[\{\log f(z_i\mid\bm{\theta})\}^2]
-\sum_{i=1}^n[\E_{\bm{\theta}\mid\bm{z};\bm{\xi}}\{\log f(z_i\mid\bm{\theta})\}]^2
\label{waic}
\end{align}
and it can be implemented in the MCMC method. In this sense, it does not seem to have any weaknesses compared with the DIC. In fact, it is included in some standard Bayesian textbooks, such as \cite{GelCSDVR13}, and has become a standard tool. 

On the other hand, there are two concerns that arise when we try to apply the WAIC to relatively complex Bayesian modeling. The first is that there is no penalty for using a class of complex prior distributions. For example, as prior distributions, let us consider $\pi_{1,n}(\bm{\theta};\xi)$ and $\pi_{2,n}(\bm{\theta};\xi_1,\xi_2)$ which includes $\pi_{1,n}(\bm{\theta};\xi)$ as a subclass. Here, we have added $n$ for the later discussion, but there is no need to worry about it at the moment. If we write the criterion for $\pi_{1,n}(\bm{\theta};\xi)$ as ${\rm WAIC}_1(\xi)$ and the criterion for $\pi_{2,n}(\bm{\theta};\xi_1,\xi_2)$ as ${\rm WAIC}_2(\xi_1,\xi_2)$, then it always holds that $\min_{\xi}{\rm WAIC}_1(\xi)>\min_{\xi_1,\xi_2}{\rm WAIC}_2(\xi_1,\xi_2)$. That is, $\pi_{1,n}(\bm{\theta};\xi)$ is never selected and $\pi_{2,n}(\bm{\theta};\xi_1,\xi_2)$ is always selected in this example. Similarly, if we consider a candidate class of prior distributions that includes $\pi_{2,n}(\bm{\theta};\xi_1,\xi_2)$, it will always be selected, and if we repeat this selection, over-fitting will occur. In particular, considerable over-fitting will be inevitable in classes of infinite dimensional prior distributions, such as the one in function space. Considering this situation, let us define $\xi$ that minimizes some criterion ${\rm IC}_1(\xi)$ for $\pi_{1,n}(\bm{\theta};\xi)$ as $\hat{\xi}_{\bm{z}}$ and define $(\xi_1,\xi_2)$ that minimizes some criterion ${\rm IC}_2(\xi_1,\xi_2)$ for $\pi_{2,n}(\bm{\theta};\xi_1,\xi_2)$ as $(\hat{\xi}_{1,\bm{z}},\hat{\xi}_{2,\bm{z}})$. We evaluate the bias caused by $\hat{\xi}_{\bm{z}}$ and the bias caused by $(\hat{\xi}_{1,\bm{z}},\hat{\xi}_{2,\bm{z}})$ and add them to ${\rm IC}_1(\hat{\xi}_{\bm{z}})$ and ${\rm IC}_2(\hat{\xi}_{1,\bm{z}},\hat{\xi}_{2,\bm{z}})$, respectively. In other words, like the Akaike Bayesian information criterion (ABIC; \citealt{Aka80}), we include penalties when determining the hyper-parameters from the data. In particular, like the AIC, the ABIC penalizes the dimension of the hyper-parameters, i.e., $1$ for $\hat{\xi}_{\bm{z}}$ and $2$ for $(\hat{\xi}_{1,\bm{z}},\hat{\xi}_{2,\bm{z}})$. In this paper, we derive a mathematically guaranteed penalty based on asymptotics.

If we do not consider classes of prior distributions with different complexities, is there still any problem in applying the WAIC? For example, suppose that we believe that $\pi_n(\bm{\theta};\xi)$ is the only class and that it is a Laplace distribution $(\xi/2)^p\exp(-\xi\|\bm{\theta}\|_1)$. In this case, the Bayesian estimator is a least absolute shrinkage and selection operator (LASSO; \citealt{Tib96}) estimator, of which $\xi=0$ corresponds to the maximum likelihood estimator. According to \cite{Wat15}, the WAIC in this case allows for selection of the prior distribution. Actually, the criterion reflects the higher order terms of the risk, but it also means that the criterion evaluates subtle differences from the maximum likelihood estimation. One of the valuable features of the LASSO is that it does the variable selection at the same time as the estimation, whereas the WAIC assesses the risk in a world around the maximum likelihood estimation without variable selection. This can be intuitively understood from the fact that when the sample size $n$ is infinite, the effect of the prior distribution, which is $\O(1)$, disappears. Although it may seem like an artificial way of thinking, we can let $n_0$ be the actual size of the data currently available and consider a prior distribution of the form $\pi_n(\bm{\theta};\xi)=\pi(\bm{\theta};\bm{\xi})^{n/n_0}$. In the example where the Laplace distribution is as $\pi(\bm{\theta};\bm{\xi})$, the prior distribution is $p\log(\xi/2)-n(\xi/n_0)\|\bm{\theta}\|_1$, which emphasizes the prior distribution in relation to the data size. It is not emphasized in the current data because $n/n_0=1$, but it is emphasized when considering asymptotics. However, this does not mean that we should assume that the prior distribution will change in the future, but rather that asymptotics is mostly used to approximate something when the sample size is relatively large, and in that case, we should consider asymptotics in which the effect of the prior distribution remains. In this paper, under this asymptotic setting, we derive the information criterion by following the classical derivation of the AIC without doing anything special. In the Laplace distribution example, the selection problem of $\xi$ does not depend on $n_0$.

Our contributions as a methodology are summarized as follows:
\begin{itemize}
\item
For the Bayesian predictive distribution when using a prior distribution that depends on the data size $n$ and keeps the effect at $n\to\infty$, we derive an asymptotically unbiased estimator of the risk, which is the source of WAIC. 
\item
To avoid overfitting that would occur with the use of WAIC when also trying to select a prior distribution, we derive the penalty for the complexity of the class of prior distributions that is to be attached to the above asymptotically unbiased estimator.
\end{itemize}
We name it the prior intensified information criterion (PIIC). The performance evaluation and by-products of this development are summarized as follows:
\begin{itemize}
\item
When the data size $n$ is not large relative to the parameter dimension $p$, comparing with the WAIC, the PIIC gives better predictions in almost all cases and in some cases significantly better and different predictions.
\item
When a prior distribution is used for sparse estimation, we show that the penalty of the PIIC depends only on the so-called active set.
\item
To explore the possibilities for a meaningful generalization of the PIIC, we extend it to the case of causal inference based on propensity scores.
\end{itemize}

Here, we would like to add a few notes. The first is about so-called Bayesian cross validations (\citealt{GelDC92}, \citealt{VehGG17}), which include a Bayesian leave-one-out (LOO) and an importance sampling (IS) LOO. While these are comparable standard criteria to the WAIC, the WAIC and Bayesian cross validations target slightly different risks and they approximate each risk with good accuracy. In this paper, we regard the former as a natural risk and therefore use the WAIC for comparison. The second is about so-called singular statistical models. The WAIC is a valid criterion for those models, and therefore one may think that it is not used for regular statistical models. However, in important examples such as \cite{Nature17}, \cite{Nature18} and \cite{Nature19}, and in textbooks such as \cite{GelCSDVR13} and \cite{KornerRVGAK15}, the WAIC appears as a tool for regular models. Considering it, this paper will focus only on regular models, and we will only mention singular models in the last section.

The organization of this paper is as follows. First, in Section \ref{sec2}, we explain the model and assumption and summarize how to derive the information criterion for Bayesian predictive distributions. Section \ref{sec3} is the main part; it introduces the intensified prior distribution. First, we derive the criterion without considering the bias caused by selecting the class of prior distributions. We will also deal with the Laplace distribution as the prior distribution, and show that the criterion can be made to depend only on the active set, thus reducing the computational load, by introducing the asymptotics of sparse estimation. At the end of this section, we derive the penalties for the complexity of the class of prior distributions and give the final form of the PIIC. In Sections \ref{sec4} and \ref{sec5}, we compare the WAIC and PIIC in numerical experiments and a real data analysis. In each section, we check under which settings the PIIC clearly excels in prediction performance and how much the two criteria differ in their estimates. In Section \ref{sec6}, to show the development potential of the PIIC, we extend it to a simple case in propensity score analysis, which makes use of the inverse probability weighted estimation. We summarize our conclusions in Section \ref{sec7}.

\section{Preparation}
\label{sec2}
\subsection{Model and assumption}
\label{sec2_1}
Letting $f(\cdot;\mid\bm{\theta})$ be a probability function with parameter $\bm{\theta}\in\mathbb{R}^p$ and $\pi_n(\cdot;\bm{\xi})$ be a prior probability function with hyper-parameter $\bm{\xi}\in\mathbb{R}^q$, we define independent random variables as follows:
\begin{align}
z_1,z_2,\ldots,z_n, \tilde{z}_1, \tilde{z}_2,\ldots,\tilde{z}_n\stackrel{\rm i.i.d.}{\sim}f(\cdot\mid\bm{\theta}), \qquad \bm{\theta}\sim\pi_n(\cdot;\bm{\xi}).
\label{model}
\end{align}
Here, $z_i$ is the data for the $i$-th sample and $\tilde{z}_i$ is the corresponding future data. In regression analysis, $z_i$ is regarded as $(y_i,\bm{x}_i)$, where $y_i$ is the objective variable and $\bm{x}_i$ is the explanatory variable vector. The WAIC, which is the subject of comparison in this paper, can handle singular statistical models, which is one of its big merits, but we will limit ourselves to regular statistical models here. Specifically, we impose the following conditions, where $\partial^+$ denotes the right partial derivative and $\partial^-$ denotes the left partial derivative:

{\centering
\begin{tabular}{lp{.88\linewidth}}
\vspace{-1mm}\\
(C1) & The parameter sets $\bm{\Theta}$ and $\bm{\Xi}$ for $\bm{\theta}$ and $\bm{\xi}$ are compact in $\mathbb{R}^p$ and $\mathbb{R}^q$, respectively.
\vspace{2mm}\\
(C2) & The functions $\log f(z\mid\bm{\theta})$ and $\log\pi_n(\bm{\theta};\bm{\xi})$ are of class C$^2$ with respect to $\bm{\theta}$ and $\bm{\xi}$, respectively, and there exists $h(z)$ such that the absolute value of each component of $\partial^2\log f(z\mid\bm{\theta})/\partial\bm{\theta}\partial\bm{\theta}'$ is less than $h(z)$ and $\E_z\{h(z)\}<\infty$.
\vspace{2mm}\\
(C3) & The parameter vector $\bm{\theta}$ is divided into sub-vectors $\bm{\theta}^{(1\cup2)}\ (\mathbb{R}^{p_{1\cup2}})$ and $\bm{\theta}^{(3)}\ (\mathbb{R}^{p_{3}})$, and accordingly $\log\pi_n(\bm{\theta};\bm{\xi})$ is also divided into $\log\pi_n^{(1\cup2)}(\bm{\theta}^{(1\cup2)};\bm{\xi})+\log\pi_n^{(3)}(\bm{\theta}^{(3)};\bm{\xi})$. With respect to the component $\theta_j$ of $\bm{\theta}^{(1\cup2)}$, $\log\pi_n^{(1\cup2)}(\bm{\theta}^{(1\cup2)};\bm{\xi})$ is continuous but not differentiable at $\theta_j=0$, and it holds that $\partial^-\log\pi_n^{(1\cup2)}(\bm{0}_{p_{1\cup2}};\bm{\xi})/\partial\theta_j\ge 0$ and $\partial^+\log\pi_n^{(1\cup2)}(\bm{0}_{p_{1\cup2}};\bm{\xi})/\partial\theta_j\le 0$. The functions $\log\pi_n^{(1\cup2)}(\bm{\theta}^{(1\cup2)};\bm{\xi})$ and $\log\pi_n^{(3)}(\bm{\theta}^{(3)};\bm{\xi})$ are of class C$^2$ with respect to $\bm{\theta}^{(1\cup2)}\setminus\tilde{\bm{\theta}}$ and $\bm{\theta}^{(3)}$, respectively. 
\vspace{2mm}\\
(C4) & The functions $f(z\mid\bm{\theta})$ and $\pi_n(\bm{\theta};\bm{\xi})$ are identifiable; i.e., if $f(z\mid\bm{\theta}^{\dagger})=f(z\mid\bm{\theta}^{\ddagger})$, $\bm{\theta}^{\dagger}=\bm{\theta}^{\ddagger}$, and if $\pi_n(\bm{\theta};\bm{\xi}^{\dagger})=\pi_n(\bm{\theta};\bm{\xi}^{\ddagger})$, $\bm {\xi}^{\dagger}=\bm{\xi}^{\ddagger}$.
\vspace{2mm}\\
(C5) & For the function $\E_z\{\log f(z\mid\bm{\theta})\}+n^{-1}\log\pi_n(\bm{\theta};\bm{\xi})$, the only maximum point for $\bm{\theta}$ for fixed $\bm{\xi}$, which is denoted by $\bm{\theta }^*_{\bm{\xi}}$, is an interior point of $\bm{\Theta}$. In addition, for the function $\E_z\{\log f(z\mid\bm{\theta}^*_{\bm{\xi}})\}$, the only maximum point for $\bm{\xi}$, which is denoted by $\bm{\xi}^*$, is an interior point of $\bm{\Xi}$.
\vspace{2mm}\\
\end{tabular}
}

\noindent
Letting $\bm{z}=(z_1,z_2,\ldots,z_n)$, we denote the posterior mean by $\E_{\bm{\theta}\mid\bm{z};\bm{\xi}}(\cdot)$, which is the expectation with respect to the posterior distribution of $\bm{\theta}$, which is $f(\bm{z}\mid\bm{\theta})\pi_n(\bm{\theta};\bm{\xi})/\int_{\bm{\theta}}f(\bm{z}\mid\bm{\theta})\pi_n(\bm{\theta};\bm{\xi}){\rm d}\bm{\theta}$. Note that the Bayesian predictive distribution $f(z\mid\bm{z};\bm{\xi})$ can be written as $\E_{\bm{\theta}\mid\bm{z};\bm{\xi}}\{f(z\mid\bm{\theta})\}$.

\subsection{Information criterion}
\label{sec2_2}
The most basic criterion for statistical model selection, AIC, is an asymptotic evaluation of the Kullback-Leibler divergence between the predictive distribution based on the maximum likelihood estimation and the true distribution minus a constant, which is $-\sum_{i=1}^n\E_{\tilde{z}_i}\{\log f(\tilde{z}_i\mid\hat {\bm{\theta}}_{\bm{z}})\}$. Here $\hat{\bm{\theta}}_{\bm{z}}$ is the maximum likelihood estimator. Specifically, the negative maximum log likelihood $-\sum_{i=1}^n\log f(z_i\mid\hat{\bm{\theta}}_{\bm{z}})$ is used as the first evaluation, and letting $b^{\rm limit}_{\rm ML}$ be the weak limit of
\begin{align*}
\sum_{i=1}^n\log f(z_i\mid\hat{\bm{\theta}}_{\bm{z}}) - \sum_{i=1}^n\log f(z_i\mid\bm{\theta}^*) - \sum_{i=1}^n\log f(\tilde{z}_i\mid\hat{\bm{\theta}}_{\bm{z}}) + \sum_{i=1}^n\log f(\tilde{z}_i\mid\bm{\theta}^*),
\end{align*}
$\E(b^{\rm limit}_{\rm ML})$ is evaluated as the asymptotic bias for the correction. Then, in regular statistical models, this asymptotic bias becomes the number of parameters, and the conventional AIC is obtained. Here, $\bm{\theta}^*$ is usually set as the limit of $\hat{\bm{\theta}}_{\bm{z}}$. A number of information criteria based on this idea have been developed since then, such as the regularization information criterion (RIC; \citealt{Shi89}) for non-sparse regularized estimation and the generalized information criterion (GIC; \citealt{KonK96}) and generalized Akaike information criterion (GAIC; \citealt{LvL14}) for more generalized estimation. 

In Bayesian statistics, the standard for a long time had been to use, for example, the Bayes factor for model selection without this idea, but \cite{SpiBCL02} caused a stir. The DIC proposed there involves evaluating the Kullback-Leibler divergence between the Bayesian predictive distribution and the true distribution, and is expressed as \eqref{dic}. The penalty term in the DIC is not as simple as the number of parameters, but it is based on a Bayesian posterior distribution, and it can be easily determined by using the MCMC method. After pointing out that the DIC is not always guaranteed by statistical asymptotics, \cite{Wat10} derived the WAIC, which is guaranteed by statistical asymptotics, including the case where the identifiability condition (C4) does not necessarily hold when $\pi_n(\bm{\theta};\bm{\xi})=\O(1)$. The WAIC is an asymptotic evaluation of $-\sum_{i=1}^n\E_{\tilde{z}_i}\{\log f(\tilde{z}_i\mid\bm{z};\bm{\xi})\}$, which is the Kullback-Leibler divergence between the Bayesian predictive distribution and the true distribution minus a constant. The negative logarithm of the Bayesian predictive distribution plugging the data, $-\sum_{i=1}^n\log f(z_i\mid\bm{z};\bm{\xi})$, is used as the initial evaluation, and the asymptotic evaluation of the expectation of 
\begin{align}
\sum_{i=1}^n\log f(z_i\mid\bm{z};\bm{\xi}) - \sum_{i=1}^n\log f(z_i\mid\bm{\theta}^*_{\bm{\xi}}) - \sum_{i=1}^n\log f(\tilde{z}_i\mid\bm{z};\bm{\xi}) + \sum_{i=1}^n\log f(\tilde{z}_i\mid\bm{\theta}^*_{\bm{\xi}})
\label{bias0}
\end{align}
is used for the correction. Here, $\bm{\theta}^*_{\bm{\xi}}$ is the limit of the Bayesian estimator of $\bm{\theta}$. Specifically, the sum of the second and third terms in \eqref{waic} is given as the asymptotic evaluation of the expectation. Similarly to the DIC, this can be implemented by using the MCMC method. Not only is the WAIC mathematically valid, but it has also been reported to outperform the DIC in numerical experiments. In this paper, the weak limit of \eqref{bias0} is defined as $b^{\rm limit}$, and $\E(b^{\rm limit})$ is used as the asymptotic bias for the correction.

\section{Main result}
\label{sec3}
\subsection{Introducing the intensified prior}
\label{sec3_0}
Letting $n_0$ be the actual data size, we assume that the prior distribution is written as
\begin{align*}
\pi_n(\bm{\theta};\bm{\xi})=\pi(\bm{\theta};\bm{\xi})^{n/n_0}.
\end{align*}
Even if we change from $n$ to $n^{\alpha}\ (0\le\alpha<1)$ in this assumption, the subsequent discussion only becomes easier.

Although we make the prior distribution depend on $n$, we do not really think that the prior distribution changes as the data size increases. Statistical asymptotics has two purposes. The first is to approximate something, such as the distribution of an estimator, when the data size actually increases in the future. The second is to approximate something based on a finite but large number of samples, without considering that the data size will actually increase in the future. An example that is easy to explain is the change-point problem, where asymptotic theory assumes that the data size increases in the same proportion before and after the change point, but of course it does not assume that this will actually happen. Most purposes would suit the latter approximation, and in fact weconsider the latter here. Incidentally, asymptotics for showing oracle properties, which often appear in the context of sparse estimation, would be the first purpose. This means that when the data size increases, we can get an asymptotic distribution of the estimator like the one from the oracle, so we will not use that distribution for, for example, constructing confidence intervals. This is because the estimators for finite samples are always shrunk to zero, and an asymptotic distribution without this effect is quite different from the distribution for finite samples.

\subsection{Evaluation of penalty for intensified prior}
\label{sec3_1}
As mentioned in Section \ref{sec2_2}, we asymptotically evaluate \eqref{bias0}. If we denote $z_i$ and $\tilde{z}_i$ by $\breve{z}$ to uniformly handle the first and third items, the logarithm is expressed as
\begin{align}
\log\int f(\breve{z}\mid\bm{\theta})f(\bm{z}\mid\bm{\theta})\pi(\bm{\theta};\bm{\xi})^{n/n_0}{\rm d}\bm{\theta}-\log\int f(\bm{z}\mid\bm{\theta})\pi(\bm{\theta};\bm{\xi})^{n/n_0}{\rm d}\bm{\theta}.
\label{logpd}
\end{align}
Applying a Laplace approximation to each of these integrals (see \citealt{TieK86}), we can see that \eqref{logpd} becomes
\begin{align*}
\log g(\bm{z},\hat{\bm{\theta}}_{\breve{z},\bm{z},\bm{\xi}};\bm{\xi})-\log g(\bm{z},\hat{\bm{\theta}}_{\bm{z},\bm{\xi}};\bm{\xi})-\frac{1}{2}\log\frac{|\hat{\bm I}_{1,\breve{z},\bm{z},\bm{\xi}}(\hat{\bm{\theta}}_{\breve{z},\bm{z},\bm{\xi}})|}{|\hat{\bm I}_{1,\bm{z},\bm{\xi}}(\hat{\bm{\theta}}_{\bm{z},\bm{\xi}})|}+\log f(\breve{z}\mid\hat{\bm{\theta}}_{\breve{z},\bm{z},\bm{\xi}})+\OP\Big(\frac{1}{n^2}\Big).
\end{align*}
Here,
\begin{align}
& \log g(z_i,\bm{\theta};\bm{\xi})\equiv\log f(z_i\mid\bm{\theta})+\frac{1}{n_0}\log\pi(\bm{\theta};\bm{\xi}), \ \log g(\bm{z},\bm{\theta};\bm{\xi})\equiv\sum_{i=1}^n\log g(z_i,\bm{\theta};\bm{\xi})
\notag \\
& \hat{\bm{\theta}}_{\breve{z},\bm{z},\bm{\xi}}\equiv\argmax_{\bm{\theta}}\{\log g(\bm{z},\bm{\theta};\bm{\xi})+\log f(\breve{z}\mid\bm{\theta})\}, \ \hat{\bm{\theta}}_{\bm{z},\bm{\xi}}\equiv\argmax_{\bm{\theta}}\{\log g(\bm{z},\bm{\theta};\bm{\xi})\}
\notag \\
& \hat{\bm{I}}_{1,\breve{z},\bm{z},\bm{\xi}}(\bm{\theta})\equiv-\frac{1}{n}\frac{\partial^2}{\partial{\bm{\theta}}\partial{\bm{\theta}}'}\{\log g(\bm{z},\bm{\theta};\bm{\xi})+\log f(\breve{z}\mid\bm{\theta})\}, \ \hat{\bm{I}}_{1,\bm{z},\bm{\xi}}(\bm{\theta})\equiv-\frac{1}{n}\frac{\partial^2}{\partial{\bm{\theta}}\partial{\bm{\theta}}'}\log g(\bm{z},\bm{\theta};\bm{\xi})
\label{HatI1}
\end{align}
are respectively the logarithms of the simultaneous probability functions of the data and parameters, the Bayesian estimators of $\bm{\theta}$, and the empirical Fisher information matrices. From this and the fact that $\hat{\bm{\theta}}_{\breve{z},\bm{z},\bm{\xi}}-\hat{\bm{\theta}}_{\bm{z},\bm{\xi}}=\OP(n^{-1})$, we have
\begin{align}
\log f(\breve{z}\mid\bm{z};\bm{\xi})
=\log f(\breve{z}\mid\hat{\bm{\theta}}_{\bm{z},\bm{\xi}})+\OP\Big(\frac{1}{n}\Big).
\label{diff}
\end{align}
In light of \eqref{diff}, \cite{KonK08} (in Section 9.4 of that paper) proposed the GIC for the Bayesian prediction method under the assumotion that the bias correction terms for the Bayesian prediction and penalized maximum likelihood estimation are the same. On the other hand, the bias correction term is actually $\OP(1)$, which is obtained from \eqref{diff} as follows: $\sum_{i=1}^n\log f(z_i\mid\bm{z};\bm{\xi})-\sum_{i=1}^n\log f(z_i\mid\hat{\bm{\theta}}_{\bm{z},\bm{\xi}})$, so the GIC cannot be said to be guaranteed by asymptotics. Consequently, if we evaluate the $\OP(n^{-1})$ term in \eqref{diff}, we find that it is
\begin{align*}
a^{\dagger}(\breve{z}\mid\bm{z},\bm{\xi})\equiv \ & \frac{1}{2}{\rm tr}\Big[\hat{\bm{I}}_{1,\bm{z},\bm{\xi}}(\hat{\bm{\theta}}_{\bm{z},\bm{\xi}})^{-1}\Big\{\frac{\partial^2}{\partial\bm{\theta}\partial\bm{\theta}'}\log f(\breve{z}\mid\hat{\bm{\theta}}_{\bm{z},\bm{\xi}})
-\frac{\partial}{\partial\bm{\theta}}\log f(\breve{z}\mid\hat{\bm{\theta}}_{\bm{z},\bm{\xi}})
\\
& \phantom{\frac{1}{2}{\rm tr}\Big[} \frac{\partial}{\partial\bm{\theta}'}\log f(\breve{z}\mid\hat{\bm{\theta}}_{\bm{z},\bm{\xi}})
- \frac{\partial}{\partial\bm{\theta}}\log f(\breve{z}\mid\hat{\bm{\theta}}_{\bm{z},\bm{\xi}})\frac{\partial}{\partial\bm{\theta}'}\log |\hat{\bm{I}}_{1,\bm{z},\bm{\xi}}(\hat{\bm{\theta}}_{\bm{z},\bm{\xi}})|\Big\}\Big]
\end{align*}
divided by $n$. Therefore, \eqref{bias0}is evaluated as
\begin{align}
& \sum_{i=1}^n\log f(z_i\mid\hat{\bm{\theta}}_{\bm{z},\bm{\xi}}) - \sum_{i=1}^n\log f(z_i\mid\bm{\theta}^*_{\bm{\xi}}) + \frac{1}{n}\sum_{i=1}^na^{\dagger}(z_i\mid\bm{z};\bm{\xi})
\notag \\
& - \sum_{i=1}^n\log f(\tilde{z}_i\mid\hat{\bm{\theta}}_{\bm{z},\bm{\xi}}) + \sum_{i=1}^n\log f(\tilde{z}_i\mid\bm{\theta}^*_{\bm{\xi}}) - \frac{1}{n}\sum_{i=1}^na^{\dagger}(\tilde{z}_i\mid\bm{z},\bm{\xi}) + \oP(1).
\label{bias1}
\end{align}
The expectations of $a^{\dagger}(z_i\mid\bm{z},\bm{\xi})$ and $a^{\dagger}(\tilde{z}_i\mid\bm{z},\bm{\xi})$ are asymptotically equivalent, and the third and sixth terms cancel out in the asymptotic evaluation of \eqref{bias1}. Then, as in the usual derivation of AIC-type information criteria, the sum of the first, second, fourth, and fifth items of \eqref{bias1} is given by Taylor expansions:
\begin{align}
& \frac{1}{n}\sum_{i,j=1}^n\Big[\frac{\partial}{\partial\bm{\theta}'}\log f(z_i\mid\bm{\theta}^*_{\bm{\xi}})-\E\Big\{\frac{\partial}{\partial\bm{\theta}'}\log f(z_i\mid\bm{\theta}^*_{\bm{\xi}})\Big\}\Big] \hat{\bm{I}}_{1,\bm{z},\bm{\xi}}(\bm{\theta}^*_{\bm{\xi}})^{-1} \frac{\partial}{\partial\bm{\theta}}\log g(z_j,\bm{\theta}^*_{\bm{\xi}};\bm{\xi})
\notag \\
& - \frac{1}{n}\sum_{i,j=1}^n\Big[\frac{\partial}{\partial\bm{\theta}'}\log f(\tilde{z}_i\mid\bm{\theta}^*_{\bm{\xi}})-\E\Big\{\frac{\partial}{\partial\bm{\theta}'}\log f(\tilde{z}_i\mid\bm{\theta}^*_{\bm{\xi}})\Big\}\Big] \hat{\bm{I}}_{1,\bm{z},\bm{\xi}}(\bm{\theta}^*_{\bm{\xi}})^{-1} \frac{\partial}{\partial\bm{\theta}}\log g(z_j,\bm{\theta}^*_{\bm{\xi}};\bm{\xi})
\notag \\
& + \oP(1).
\label{taylor}
\end{align}
Note that the quadratic terms in the Taylor expansions cancel each other out in the first and second expansions. In the first sum when $i\neq j$ and in the second sum, the term for $j$ is independent of the term for $i$ and has expectation $0$. Thus, letting $\bm{u}_1$, $\bm{u}_2$ and $\bm{u}_3$ be independent random vectors distributed according to ${\rm N}(\bm{0},\bm{I}_{2,\bm{\xi}}(\bm{\theta}^*_{\bm{\xi}}))$, we obtain
\begin{align*}
b^{\rm limit} = {\rm tr}\big\{\bm{I}_{1,\bm{\xi}}^{-1}\big(\bm{\theta}^*_{\bm{\xi}}\big)\bm{I}_{2,\bm{\xi}}\big(\bm{\theta}^*_{\bm{\xi}}\big)\big\} + \bm{u}_1'\bm{I}_{1,\bm{\xi}}^{-1}\big(\bm{\theta}^*_{\bm{\xi}}\big)\bm{u}_3 - \bm{u}_2'\bm{I}_{1,\bm{\xi}}^{-1}\big(\bm{\theta}^*_{\bm{\xi}}\big)\bm{u}_3,
\end{align*}
where 
\begin{align}
\bm{I}_{1,\bm{\xi}}(\bm{\theta}) \equiv \E_{\breve{z}}\Big\{-\frac{\partial^2}{\partial\bm{\theta}\partial\bm{\theta}'}\log g(\breve{z},\bm{\theta};\bm{\xi})\Big\}, \ \bm{I}_{2,\bm{\xi}}(\bm{\theta}) \equiv \E_{\breve{z}}\Big\{\frac{\partial}{\partial\bm{\theta}}\log g(\breve{z},\bm{\theta};\bm{\xi}) \frac{\partial}{\partial\bm{\theta}'}\log g(\breve{z},\bm{\theta};\bm{\xi})\Big\}.
\label{Idef}
\end{align}
Taking the expectation, the second and third terms on the right-hand side become $0$, which leads us to the following theorem:

\begin{theorem}
Define $\bm{I}_{1,\bm{\xi}}(\bm{\theta})$ and $\bm{I}_{2,\bm{\xi}}(\bm{\theta})$ in \eqref{Idef}, and let $\pi_n(\bm{\theta};\bm{\xi})$ be $\O(n)$. Then, the asymptotic bias of $-\sum_{i=1}^n\log f(z_i\mid\bm{z};\bm{\xi})$ using the method described in Section \ref{sec2_2} is given by ${\rm tr}\{\bm{I}_{1,\bm{\xi}}(\bm{\theta}^*_{\bm{\xi}})^{-1}\allowbreak\bm{I}_{2,\bm{\xi}}(\bm{\theta}^*_{\bm{\xi}})\}$.
\label{th1}
\end{theorem}

\noindent 
In light of this theorem and the definition of \eqref{HatI1}, we prepare
\begin{align*}
\hat{\bm{I}}_{2,\bm{z},\bm{\xi}}(\bm{\theta})\equiv\frac{1}{n}\sum_{i=1}^n\frac{\partial}{\partial\bm{\theta}}\log g(z_i,\bm{\theta};\bm{\xi})\frac{\partial}{\partial\bm{\theta}'}\log g(z_i,\bm{\theta};\bm{\xi}).
\end{align*}
Then, as the prior intensified information criterion, we propose
\begin{align*}
{\rm PIIC} \equiv -\sum_{i=1}^n\log f(z_i\mid\bm{z};\bm{\xi}) + {\rm tr}\{\hat{\bm{I}}_{1,\bm{z},\bm{\xi}}(\hat{\bm{\theta}}_{\bm{z},\bm{\xi}})^{-1} \hat{\bm{I}}_{2,\bm{z},\bm{\xi}}(\hat{\bm{\theta}}_{\bm{z},\bm{\xi}})\}.
\end{align*}
That is, we can say that the information criterion given in Section 9.4 of \cite{KonK08} is valid, which has been found as a result. Having evaluated the $\OP(n^{-1})$ term in \eqref{diff}, we can also devise an information criterion based on the log-likelihood by substituting the Bayes estimator $\hat{\bm{\theta}}_{\bm{z},\bm{\xi}}$ into $\bm{\theta}$, as follows:
\begin{align*}
-\sum_{i=1}^n\log f(z_i\mid\hat{\bm{\theta}}_{\bm{z},\bm{\xi}}) + {\rm tr}\{\hat{\bm{I}}_{1,\bm{z},\bm{\xi}}(\hat{\bm{\theta}}_{\bm{z},\bm{\xi}})^{-1} \hat{\bm{I}}_{2,\bm{z},\bm{\xi}}(\hat{\bm{\theta}}_{\bm{z},\bm{\xi}})\} + \frac{1}{n}\sum_{i=1}^na^{\dagger}(z_i\mid\bm{z};\bm{\xi}).
\end{align*}
Note that the same cannot be said from the discussion in \cite{KonK08}. 

\subsection{Combining sparse estimation}
\label{sec3_2}
As for the prior distribution in \eqref{model}, we will suppose one that leads to a sparse estimation for some parameters $\bm{\theta}^{(1\cup2)}$ in condition (C3). The most basic one is
\begin{align*}
\pi_n^{(1\cup2)}(\bm{\theta}^{(1\cup2)};\bm{\xi}) \propto \exp(-\xi\|\bm{\theta}^{(1\cup2)}\|_1)^{n/n_0},
\end{align*}
which leads to so-called the LASSO estimation (\citealt{Tib96}). Other sparse methods including the smoothly clipped absolute deviation (SCAD; \citealt{FanL01}) and minimax concave plus (MCP; \citealt{Zha10}), which are improved versions of the LASSO, and group LASSO (\citealt{YuaL06}), which is a generalized version of the LASSO, are the subject of this subsection. Also for this type of prior distribution with points indifferentiable with respect to $\bm{\theta}$, we want to use the Laplace approximation to evaluate the predictive distribution. On the other hand, the function $\log g(\bm{z},\bm{\theta};\bm{\xi})$ treated in \cite{TieK86} is assumed to be differentiable with respect to all $\bm{\theta}$'s. For the prior distribution treated in this subsection, however, the Laplace approximations can be applied without difficulty, and hereafter we will check this. For simplicity in this subsection, we will abbreviate $\bm{\xi}$ until stating Theorem \ref{th2}.

First, we will briefly describe the behavior of $\hat{\bm{\theta}}_{\breve{z},\bm{z}}$ and $\hat{\bm{\theta}}_{\bm{z}}$ in this setting. From the previous subsection, we know that they converge in probability to $\bm{\theta}^*$. This vector $\bm{\theta}^*$ is sparse, that is, some of the components are exactly zero. From now on, we will write $\bm{\theta}^{*(1)}$ as the sub-vector of $\bm{\theta}^*$ from which the zero components are extracted, and write $\bm{\theta}^{*(2\cup 3)}$ as the sub-vector of $\bm{\theta}^*$ from which the non-zero components are extracted. That is, when $p_1$ is the number of components that are zero, $\bm{\theta}^{*(1)}=\bm{0}_{p_1}$. In addition, we denote the sub-vectors of $\hat{\bm{\theta}}_{\bm{z}}$ corresponding to $\bm{\theta}^{*(1)}$ and $\bm{\theta}^{*(2\cup 3)}$ by $\hat{\bm{\theta}}_{\bm{z}}^{(1)}$ and $\hat{\bm{\theta}}_{\bm{z}}^{(2\cup 3)}$, respectively. As we will see later, the convergence of $\hat{\bm{\theta}}_{\bm{z}}^{(1)}$ to $\bm{0}_{p_1}$ is known to be fast. On the other hand, asymptotic normality is known to hold for $\hat{\bm{\theta}}_{\bm{z}}^{(2\cup 3)}$, just like the usual estimator.

Next, let us approximate $\log\{f(\breve{z}\mid\bm{\theta})g(\bm{z},\bm{\theta})\}$ around $\bm{\theta}=\hat{\bm{\theta}}_{\breve{z},\bm{z}}$ and $\log g(\bm{z },\bm{\theta})$ around $\bm{\theta}=\hat{\bm{\theta}}_{\bm{z}}$. Although $\hat{\bm{\theta}}_{\breve{z},\bm{z}}$ and $\hat{\bm{\theta}}_{\bm{z}}$ are sparse and the Taylor expansion cannot be used for, for example, $\log g(\bm{z},\bm{\theta})$, we have only to use
\begin{align*}
& \log g(\bm{z},\hat{\bm{\theta}}_{\bm{z}}) + 
\sum_{j\in{\cal J}^{(1)}}\big\{\theta_j^+s_{\bm{z},j}^+(\hat{\bm{\theta}}_{\bm{z}})+\theta_j^-s_{\bm{z},j}^-(\hat{\bm{\theta}}_{\bm{z}})\big\} - \frac{n}{2}(\bm{\theta}-\hat{\bm{\theta}}_{\bm{z}})' \hat{\bm{I}}_{1,\bm{z}}(\hat{\bm{\theta}}_{\bm{z}}) (\bm{\theta}-\hat{\bm{\theta}}_{\bm{z}}),
\end{align*}
where ${\cal J}^{(1)}\equiv\{j: \hat{\theta}_{\bm{z},j}=0\}$, $\theta_j^+\equiv\max(\theta_j,0)$, $\theta_j^-\equiv\min(\theta_j,0)$, $s _{\bm{z},j}^+(\hat{\bm{\theta}}_{\bm{z}})\equiv\partial^+\log g(\bm{z},\hat{\bm{\theta}}_{\bm{z}})/\allowbreak\partial\theta_j$, and $s_{\bm{z},j}^-(\hat{\bm{\theta}}_{\bm{z}})\equiv\partial^-\log g(\bm{z},\hat{\bm{\theta}}_{\bm{z}})/\partial\theta_j$. Note that the limit of $s_{\bm{z},j}^+(\hat{\bm{\theta}}_{\bm{z}})/n$ is negative and the limit of $s_{\bm{z},j}^-(\hat{\bm{\theta}}_{\bm{z}})/n$ is positive. Thus, the following asymptotic approximation is obtained:
\begin{align*}
& \int\exp\{\log g(\bm{z},\bm{\theta})\}{\rm d}\bm{\theta}
\\
& = \int \exp\Bigg[\sum_{j\in{\cal J}^{(1)}}\{
\theta_j^+s_{\bm{z},j}^+(\hat{\bm{\theta}}_{\bm{z}})+\theta_j^-s_{\bm{z},j}^-(\hat{\bm{\theta}}_{\bm{z}})\}
\\
& \phantom{= \int} \ - \frac{n}{2}(\bm{\theta}^{(2\cup 3)}-\hat{\bm{\theta}}_{\bm{z}}^{(2\cup 3)})' \hat{\bm{I}}_{1,\bm{z}}^{(2\cup 3)}(\hat{\bm{\theta}}_{\bm{z}}) (\bm{\theta}^{(2\cup 3)}-\hat{\bm{\theta}}_{\bm{z}}^{(2\cup 3)})\Bigg]{\rm d}\bm{\theta}\ g(\bm{z},\hat{\bm{\theta}}_{\bm{z}}) \Big\{1+\frac{c}{n}+\OP\Big(\frac{1}{n^2}\Big)\Big\}
\\
& = \frac{1}{n^{|{\cal J}^{(1)}|}} \big|2\pi n\hat{\bm{I}}_{1,\bm{z}}^{(2\cup 3)}(\hat{\bm{\theta}}_{\bm{z}})\big|^{1/2} 
\sum_{{\cal J}\subset{\cal J}^{(1)}} \Bigg\{ \prod_{j\in{\cal J}} \frac{1}{s_{\bm{z},j}^+(\hat{\bm{\theta}}_{\bm{z}})} \prod_{j\in{\cal J}^c} \frac{1}{s_{\bm{z},j}^-(\hat{\bm{\theta}}_{\bm{z}})} \Bigg\} 
\\
& \phantom{=} \ g(\bm{z},\hat{\bm{\theta}}_{\bm{z}}) \Big\{1+\frac{c}{n}+\OP\Big(\frac{1}{n^2}\Big)\Big\}.
\end{align*}
Here, $\bm{\theta}^{(2\cup 3)}$ is the sub-vector of $\bm{\theta}$ corresponding to $\bm{\theta}^{*(2\cup 3)}$, $\hat{\bm{I}}_{1,\bm{z}}^{(2\cup 3)}(\hat{\bm{\theta}}_{\bm{z}})$ is the sub-matrix of $\hat{\bm{I}}_{1,\bm{z}}(\hat{\bm{\theta}}_{\bm{z}})$ corresponding to $\bm{\theta}^{*(2\cup 3)}$, $c$ is a constant, and ${\cal J}^c\equiv{\cal J}^{(1)}\setminus{\cal J}$. Almost the same asymptotic approximation can be obtained for $\int\exp\{\log g(\breve{z},\bm{z},\bm{\theta})\}{\rm d}\bm{\theta}$. Because the difference between $\log g(\bm{z},\bm{\theta})$ and $\log g(\breve{z},\bm{z},\bm{\theta})$ is $\OP(1)$, we can see that the constant $c$ is common. Thus, by replacing $g(\bm{z},\bm{\theta})$ with $g(\breve{z},\bm{z},\bm{\theta})$ and defining $\hat{\bm{I}}_{1,\breve{z},\bm{z}}^{(2\cup 3)}(\hat{\bm{\theta}}_{\breve{z},\bm{z}})$, $s_{\breve{z},\bm{z},j}^+(\hat{\bm{\theta}}_{\breve{z},\bm{z}})$, and $s_{\breve{z},\bm{z},j}^-(\hat{\bm{\theta}}_{\breve{z},\bm{z}})$ in the same way, we have
\begin{align}
f(\breve{z}\mid\bm{z}) & = \frac{\int g(\breve{z},\bm{z},\bm{\theta}){\rm d}\bm{\theta}}{\int g(\bm{z},\bm{\theta}){\rm d}\bm{\theta}}
\notag \\
& = \frac{|\hat{\bm{I}}_{1,\breve{z},\bm{z}}^{(2\cup 3)}(\hat{\bm{\theta}}_{\breve{z},\bm{z}})|^{1/2}}{|\hat{\bm{I}}_{1,\bm{z}}^{(2\cup 3)}(\hat{\bm{\theta}}_{\bm{z}})|^{1/2}} \frac{\sum_{{\cal J}\subset{\cal J}^{(1)}} \{ \prod_{j\in{\cal J}} s_{\breve{z},\bm{z},j}^+(\hat{\bm{\theta}}_{\breve{z},\bm{z}})^{-1} \prod_{j\in{\cal J}^c} s_{\breve{z},\bm{z},j}^-(\hat{\bm{\theta}}_{\breve{z},\bm{z}})^{-1} \} }{\sum_{{\cal J}\subset{\cal J}^{(1)}} \{ \prod_{j\in{\cal J}} s_{\bm{z},j}^+(\hat{\bm{\theta}}_{\bm{z}})^{-1} \prod_{j\in{\cal J}^c} s_{\bm{z},j}^-(\hat{\bm{\theta}}_{\bm{z}})^{-1} \} }
\notag \\
& \ \phantom{=} \frac{g(\bm{z},\hat{\bm{\theta}}_{\breve{z},\bm{z}})}{g(\bm{z},\hat{\bm{\theta}}_{\bm{z}})} f(\breve{z}\mid\hat{\bm{\theta}}_{\breve{z},\bm{z}}) \Big\{1+\OP\Big(\frac{1}{n^2}\Big)\Big\}.
\label{spapprox}
\end{align}
Note that the $c/n$ terms have canceled out.

The same argument as in Section \ref{sec3_1} can be developed using \eqref{spapprox}, but to give a simpler and more stable criterion, we will actively incorporate the fact that it is sparse. Specifically, since $\hat{\bm{\theta}}^{(1)}_{\breve{z},\bm{z},j}$ and $\hat{\bm{\theta}}^{(1)}_{\bm{z},j}$ converge quickly to $\bm{0}_{p_1}$, we can asymptotically ignore terms related to them. As a preliminary result, letting the matrices of $\bm{I}_{1}(\bm{\theta}^*)$ and $\bm{I}_{2}(\bm{\theta}^*)$ with the extracted components corresponding to $\bm{\theta}^{(2)}$ and $\bm{\theta}^{(3)}$ be $\bm{I}^{(2\cup3)}_{1}(\bm{\theta}^*)$ and $\bm{I}^{(2\cup3)}_{2}(\bm{\theta}^*)$, respectively, the asymptotic bias of $\sum_{i=1}^n\log f(z_i\mid\bm{z})$ obtained by the method in Section \ref{sec2_2} is given by ${\rm tr}\{\bm{I}_{1}^{(2\cup3)}(\bm{\theta}^*)^{-1}\bm{I}_{2}^{(2\cup3)}(\bm{\theta}^*)\}$. To verify this, we use \eqref{spapprox} in \eqref{bias0} and then obtain
\begin{align}
& \sum_{i=1}^n \log\Bigg[ \frac{|\hat{\bm{I}}_{1,\tilde{z}_i,\bm{z}}^{(2\cup 3)}(\hat{\bm{\theta}}_{\tilde{z}_i,\bm{z}})|^{1/2}}{|\hat{\bm{I}}_{1,\bm{z}}^{(2\cup 3)}(\hat{\bm{\theta}}_{\bm{z}})|^{1/2}} \frac{\sum_{{\cal J}\subset{\cal J}^{(1)}} \{ \prod_{j\in{\cal J}} s_{\bm{z},j}^+(\hat{\bm{\theta}}_{\bm{z}}) \prod_{j\in{\cal J}^c} s_{\bm{z},j}^-(\hat{\bm{\theta}}_{\bm{z}}) \} }{\sum_{{\cal J}\subset{\cal J}^{(1)}} \{ \prod_{j\in{\cal J}} s_{\tilde{z}_i,\bm{z},j}^+(\hat{\bm{\theta}}_{\tilde{z}_i,\bm{z}}) \prod_{j\in{\cal J}^c} s_{\tilde{z}_i,\bm{z},j}^-(\hat{\bm{\theta}}_{\tilde{z}_i,\bm{z}}) \} } \frac{g(\bm{z},\hat{\bm{\theta}}_{\bm{z}})}{g(\bm{z},\hat{\bm{\theta}}_{\tilde{z}_i,\bm{z}})}\Bigg]
\notag \\
& - \sum_{i=1}^n \log\Bigg[ \frac{|\hat{\bm{I}}_{1,{z}_i,\bm{z}}^{(2\cup 3)}(\hat{\bm{\theta}}_{{z}_i,\bm{z}})|^{1/2}}{|\hat{\bm{I}}_{1,\bm{z}}^{(2\cup 3)}(\hat{\bm{\theta}}_{\bm{z}})|^{1/2}} \frac{\sum_{{\cal J}\subset{\cal J}^{(1)}} \{ \prod_{j\in{\cal J}} s_{\bm{z},j}^+(\hat{\bm{\theta}}_{\bm{z}}) \prod_{j\in{\cal J}^c} s_{\bm{z},j}^-(\hat{\bm{\theta}}_{\bm{z}}) \} }{\sum_{{\cal J}\subset{\cal J}^{(1)}} \{ \prod_{j\in{\cal J}} s_{{z}_i,\bm{z},j}^+(\hat{\bm{\theta}}_{{z}_i,\bm{z}}) \prod_{j\in{\cal J}^c} s_{{z}_i,\bm{z},j}^-(\hat{\bm{\theta}}_{{z}_i,\bm{z}}) \} } \frac{g(\bm{z},\hat{\bm{\theta}}_{\bm{z}})}{g(\bm{z},\hat{\bm{\theta}}_{{z}_i,\bm{z}})}\Bigg]
\notag \\
& +\sum_{i=1}^n\{\log f(z_i\mid\hat{\bm{\theta}}_{z_i,\bm{z}})-\log f(z_i\mid\bm{\theta}^*)\}
-\sum_{i=1}^n\{\log f(\tilde{z}_i\mid\hat{\bm{\theta}}_{\tilde{z}_i,\bm{z}})-\log f(\tilde{z}_i\mid\bm{\theta}^*)\}
+\oP(1).
\label{spapprox2}
\end{align}
Similarly to Section \ref{sec3_1}, using a function $a^{\ddagger}$, the main term of the first sum can be written as $\sum_{i=1}^na^{\ddagger}(z_i\mid\bm{z})/n$, and the main term of the second sum as $\sum_{i=1}^na^{\ddagger}(\tilde{z}_i\mid\bm{z})/n$. Since the expectations of $a^{\ddagger}(z_i\mid\bm{z})$ and $a^{\ddagger}(\tilde{z}_i\mid\bm{z})$ are asymptotically equivalent, we can see that the difference between the first and second sums in \eqref{spapprox2} is $\oP(1)$. From \cite{KniF00} and \cite{NinK16}, we have
\begin{align*}
n\hat{\bm{\theta}}_{\bm{z}}^{(1)}=\oP(1), \ \sqrt{n}(\hat{\bm{\theta}}_{\bm{z}}^{(2\cup3)}-\bm{\theta}^{*(2\cup3)})=\bm{I}^{(2\cup3)}_1(\bm{\theta}^*)^{-1} \frac{1}{\sqrt{n}}\sum_{i=1}^n\frac{\partial}{\partial\bm{\theta}^{(2\cup3)}}\log g(z_i,\bm{\theta}^*)+\oP(1),
\end{align*}
and using these, the third and fourth sums in \eqref{spapprox2} can be written as
\begin{align*}
& \frac{1}{n}\sum_{i,j=1}^n\frac{\partial}{\partial\bm{\theta}^{(2\cup3)}{}'}\log f(z_i\mid\bm{\theta}^*) \bm{I}_1^{(2\cup3)}(\bm{\theta}^*)^{-1} \frac{\partial}{\partial\bm{\theta}^{(2\cup3)}}\log g(z_j,\bm{\theta}^*)
\notag \\
& - \frac{1}{n}\sum_{i,j=1}^n\frac{\partial}{\partial\bm{\theta}^{(2\cup3)}{}'}\log f(\tilde{z}_i\mid\bm{\theta}^*) \bm{I}_1^{(2\cup3)}(\bm{\theta}^*)^{-1} \frac{\partial}{\partial\bm{\theta}^{(2\cup3)}}\log g(z_j,\bm{\theta}^*) + \oP(1)
\end{align*}
similarly to \eqref{taylor}. Since the first sum for the case of $i\neq j$ and the second sum converge in distribution to statistics with expectation zero, by writing $\bm{\xi}$ again, it can be seen that the following theorem holds:
\begin{theorem}
Define $\bm{I}_{1,\bm{\xi}}(\bm{\theta})$ and $\bm{I}_{2,\bm{\xi}}(\bm{\theta})$ in \eqref{Idef}, and let $\pi_n(\bm{\theta};\bm{\xi})$ be $\O(n)$. Then, the asymptotic bias of $-\sum_{i=1}^n\log f(z_i\mid\bm{z};\bm{\xi})$ using the method described in Section \ref{sec2_2} is given by ${\rm tr}\{\bm{I}_{1}^{(2\cup3)}(\bm{\theta}^*_{\bm{\xi}})^{-1}\bm{I}_{1}^{(2\cup3)}(\bm{\theta}^*_{\bm{\xi}})\}$.
\label{th2}
\end{theorem}

Denoting by $\hat{\bm{I}}_{1,\bm{z},\bm{\xi}}^{(2\cup3)}(\bm{\theta})$ and $\hat{\bm{I}}_{2,\bm{z},\bm{\xi}}^{(2\cup3)}(\bm{\theta})$ the sub-matrices corresponding to the active set in $\hat{\bm{I}}_{1,\bm{z},\bm{\xi}}(\bm{\theta})$ and $\hat{\bm{I}}_{2,\bm{z},\bm{\xi}}(\bm{\theta})$, where the active set is a collection of $j$'s such that $\hat{\theta}_{j,\bm{z},\bm{\xi}}\neq 0$, we find that consistent estimators of $\bm{I}_{1}^{(2\cup3)}(\bm{\theta}^*_{\bm{\xi}})$ and $\bm{I}_{2}^{(2\cup3)}(\bm{\theta}^*_{\bm{\xi}})$ are given by $\hat{\bm{I}}_{1,\bm{z},\bm{\xi}}^{(2\cup3)}(\hat{\bm{\theta}}_{\bm{z},\bm{\xi}})$ and $\hat{\bm{I}}_{2,\bm{z},\bm {\xi}}^{(2\cup3)}(\hat{\bm{\theta}}_{\bm{z},\bm{\xi}})$, respectively. In addition, by writing the predictive distribution constructed using only $\bm{\theta}$ corresponding to the active set as $f^{(2\cup 3)}(\breve{z}\mid\bm{z};\bm{\xi})$, we can easily see, in the same way as in Theorem \ref{th2}, that $\log f^{(2\cup 3)}(\breve{z}\mid\bm{z};\bm{\xi})=\log f(\breve{z}\mid\bm{z};\bm{\xi})+\oP(n^{-1})$. Thus, as the prior intensified information criterion, we propose
\begin{align*}
{\rm PIIC} = -\sum_{i=1}^n\log f^{(2\cup 3)}(z_i\mid\bm{z};\bm{\xi}) + {\rm tr}\big\{\hat{\bm{I}}_{1,\bm{z},\bm{\xi}}^{(2\cup3)}\big(\hat{\bm{\theta}}_{\bm{z},\bm{\xi}}\big)^{-1} \hat{\bm{I}}_{2,\bm{z},\bm{\xi}}^{(2\cup3)}\big(\hat{\bm{\theta}}_{\bm{z},\bm{\xi}}\big)\big\}.
\end{align*}

\subsection{Penalty evaluation for hyper-parameters in prior distribution}
\label{sec3_3}
The information criterion in Section \ref{sec3_1} or \ref{sec3_2} should take into account that the estimator is distorted by the prior distribution more than the WAIC is. On the other hand, the hyper-parameter $\bm{\xi}$ that minimizes the information criterion, $\hat{\bm{\xi}}_{\bm{z}}$, is used, but if so, the more one increases the dimension of $\bm{\xi}$, the more the increased dimension is selected, leading to overfitting. Therefore, we will target $\sum_{i=1}^n\E_{\tilde{z}_i}\{\log f(\tilde{z}_i\mid\bm{z};\hat{\bm{\xi}}_{\bm{z}})\}$ and hereafter evaluate the quantity made by replacing $\bm{\xi}$ with $\hat{\bm{\xi}}_{\bm{z}}$ in \eqref{bias0}.

The minimizer of the information criterion $\hat{\bm{\xi}}_{\bm{z}}$ asymptotically maximizes $(1/n)\sum_{i=1}^n\log \allowbreak f(z_i\mid\bm{z};\bm{\xi}) \approx (1/n)\sum_{i=1}^n\log f(z_i\mid\hat {\bm{\theta}}_{\bm{\xi}}) \approx (1/n)\sum_{i=1}^n\log f(z_i\bm{\theta}^*_{\bm{\xi}})$. On the other hand, $\bm{\xi}^*$ in condition (C4) is a parameter that maximizes $\E_{z}\{\log f(z\mid\bm{\theta}^*_{\bm{\xi}})\}$. Then, from the similarity between $(1/n)\sum_{i=1}^n\log f(z_i\mid\bm{\theta}^*_{\bm{\xi}})$ and $\E_{z}\{\log f(z\mid\bm{\theta}^*_{\bm{\xi}})\}$, we can see that $\hat{\bm{\xi}}_{\bm{z}}-\bm{\xi}^*=\oP(1)$ under the regularity condition mentioned in Section \ref{sec2_1}. Using this fact and expanding the right-hand side of $\OP(1/n)=(1/n)\sum_{i=1}^n\partial\log f(z_i\mid\bm{z};\hat{\bm{\xi}}_{\bm{z}})/\partial\bm{\xi}$ around $\hat{\bm{\xi}}_{\bm{z}}=\bm{\xi}^*$, as well as the conventional statistical asymptotics, we obtain
\begin{align*}
\hat{\bm{\xi}}_{\bm{z}}-\bm{\xi}^* = \bm{J}_{1}(\bm{\xi}^*)^{-1}\frac{1}{n}\sum_{i=1}^n\Big[\frac{\partial}{\partial\bm{\xi}}\log f(z_i\mid\bm{z},\bm{\xi}^*) - \E_{z_i}\Big\{\frac{\partial}{\partial\bm{\xi}}\log f(z_i\mid\bm{z},\bm{\xi}^*)\Big\}\Big] + \oP(n^{-1/2}),
\end{align*}
where
\begin{align}
& \bm{J}_1(\bm{\xi}) \equiv \E_{\breve{z},\bm{z}}\Big\{-\frac{\partial^2}{\partial\bm{\xi}\partial\bm{\xi}'}\log f(\breve{z}\mid\bm{z},\bm{\xi}^*)\Big\},
\notag \\
& \bm{J}_2(\bm{\xi}) \equiv \E_{\breve{z},\bm{z}}\Big\{\frac{\partial}{\partial\bm{\xi}}\log f(\breve{z}\mid\bm{z},\bm{\xi}^*) \frac{\partial}{\partial\bm{\xi}'}\log f(\breve{z}\mid\bm{z},\bm{\xi}^*)\Big\}.
\label{Jdef}
\end{align}
The matrix $\bm{J}_2(\bm{\xi})$ is defined here as it will be used later. Using a Taylor expansion, except for the quantity made by substituting $\bm{\xi}^*$ into $\bm{\xi}$ in \eqref{bias0}, the target can be evaluated by 
\begin{align*}
\sum_{i=1}^n\frac{\partial}{\partial\bm{\xi}'}\log f(z_i\mid\bm{z},\bm{\xi}^*)(\hat{\bm{\xi}}_{\bm{z}}-\bm{\xi}^*) - \sum_{i=1}^n\frac{\partial}{\partial\bm{\xi}'}\log f(\tilde{z}_i\mid\bm{z},\bm{\xi}^*)(\hat{\bm{\xi}}_{\bm{z}}-\bm{\xi}^*)
\end{align*}
plus $\oP(1)$, that is,
\begin{align}
& \sum_{i,j=1}^n\frac{\partial}{\partial\bm{\xi}'}\log f(z_j\mid\bm{z}_{-i},\bm{\xi}^*) \bm{J}_{1}(\bm{\xi}^*)^{-1} \Big[\frac{\partial}{\partial\bm{\xi}}\log f(z_i\mid\bm{z}_{-j},\bm{\xi}^*) - \E_{z_i}\Big\{\frac{\partial}{\partial\bm{\xi}}\log f(z_i\mid\bm{z}_{-j},\bm{\xi}^*)\Big\}\Big]
\notag \\
& - \sum_{i,j=1}^n\frac{\partial}{\partial\bm{\xi}'}\log f(\tilde{z}_j\mid\bm{z}_{-i},\bm{\xi}^*) \bm{J}_{1}(\bm{\xi}^*)^{-1} \Big[\frac{\partial}{\partial\bm{\xi}}\log f(z_i\mid\bm{z},\bm{\xi}^*) - \E_{z_i}\Big\{\frac{\partial}{\partial\bm{\xi}}\log f(z_i\mid\bm{z},\bm{\xi}^*)\Big\}\Big]
\label{bias3}
\end{align}
divided by $n$ plus $\oP(1)$. The matrix $\bm{J}_2(\bm{\xi})$ will be used later. Using a Taylor expansion, except for the quantity made by substituting $\bm{\xi}^*$ into $\bm{\xi}$ in \eqref{bias0}, the target can be evaluated by 
\begin{align*}
\sum_{i=1}^n\frac{\partial}{\partial\bm{\xi}'}\log f(z_i\mid\bm{z},\bm{\xi}^*)(\hat{\bm{\xi}}_{\bm{z}}-\bm{\xi}^*) - \sum_{i=1}^n\frac{\partial}{\partial\bm{\xi}'}\log f(\tilde{z}_i\mid\bm{z},\bm{\xi}^*)(\hat{\bm{\xi}}_{\bm{z}}-\bm{\xi}^*)
\end{align*}
plus $\oP(1)$, i.e.
\begin{align}
& \sum_{i,j=1}^n\frac{\partial}{\partial\bm{\xi}'}\log f(z_j\mid\bm{z}_{-i},\bm{\xi}^*) \bm{J}_{1}(\bm{\xi}^*)^{-1} \Big[\frac{\partial}{\partial\bm{\xi}}\log f(z_i\mid\bm{z}_{-j},\bm{\xi}^*) - \E_{z_i}\Big\{\frac{\partial}{\partial\bm{\xi}}\log f(z_i\mid\bm{z}_{-j},\bm{\xi}^*)\Big\}\Big]
\notag \\
& - \sum_{i,j=1}^n\frac{\partial}{\partial\bm{\xi}'}\log f(\tilde{z}_j\mid\bm{z}_{-i},\bm{\xi}^*) \bm{J}_{1}(\bm{\xi}^*)^{-1} \Big[\frac{\partial}{\partial\bm{\xi}}\log f(z_i\mid\bm{z},\bm{\xi}^*) - \E_{z_i}\Big\{\frac{\partial}{\partial\bm{\xi}}\log f(z_i\mid\bm{z},\bm{\xi}^*)\Big\}\Big]
\label{bias3}
\end{align}
divided by $n$ plus $\oP(1)$. Here, $\bm{z}_{-i}$ is $\bm{z}$ except for $z_i$ and $\bm{z}_{-j}$ is $\bm{z}$ except for $z_j$. Also note that the second-order terms in the Taylor expansion cancel each other out. On the basis of the bias evaluation described in Section \ref{sec2_2}, we define the weak limit of this as $b^{\rm limit}$. In \eqref{bias3}, the terms for $i$ and for $j$ are uncorrelated in the first sum for the case of $i\neq j$ and are uncorrelated in the second sum in every case. To be precise, the main terms are made to be uncorrelated by removing $z_i$ or $z_j$ from $\bm{z}$. Accordingly, letting $\bm{v}_1$, $\bm{v}_2$ and $\bm{v}_3$ be independent random vectors distributed according to ${\rm N}(\bm{0},\bm{J}_2(\bm{\xi}^*))$, we obtain
\begin{align*}
b^{\rm limit} = {\rm tr}\{\bm{J}_1(\bm{\xi}^*)^{-1}\bm{J}_2(\bm{\xi}^*)\} + \bm{v}_1'\bm{J}_1(\bm{\xi}^*)^{-1}\bm{v}_3 - \bm{v}_2'\bm{J}_1(\bm{\xi}^*)^{-1}\bm{v}_3.
\end{align*}
Taking the expectations, the second and third terms on the right-hand side become zero, and the following theorem holds:

\begin{theorem}
Define $\bm{I}_{1,\bm{\xi}}(\bm{\theta})$ and $\bm{I}_{2,\bm{\xi}}(\bm{\theta})$ in \eqref{Idef}, define $\bm{J}_1(\bm{\xi})$ and $\bm{J}_2(\bm{\xi})$ by \eqref{Jdef}, and let $\pi_n(\bm{\theta};\bm{\xi})$ be $\O(n)$. Then, the asymptotic bias of $-\sum_{i=1}^n\log f(z_i\mid\bm{z},\hat{\bm{\xi}}_{\bm{z}})$ using the method described in Section \ref{sec2_2} is given by ${\rm tr}\{\bm{I}_{1,\bm{\xi}^*}(\bm{\theta}^*_{\bm{\xi}^*})^{-1}\bm{I}_{2,\bm{\xi}^*}(\bm{\theta}^*_{\bm{\xi}^*})\}+{\rm tr}\{\bm{J}_1(\bm{\xi}^*)^{-1}\bm{J}_2(\bm{\xi}^*)\}$.
\label{th3}
\end{theorem}

\noindent
From this theorem, we define
\begin{align*}
& \hat{\bm{J}}_{1,\bm{z}}(\bm{\xi})\equiv -\frac{1}{n}\sum_{i=1}^n\frac{\partial^2}{\partial\bm{\xi}\partial\bm{\xi}'}\log f(z_i\mid\bm{z},\bm{\xi}),
\\
& \hat{\bm{J}}_{2,\bm{z}}(\bm{\xi})\equiv\frac{1}{n}\sum_{i=1}^n\frac{\partial}{\partial\bm{\xi}}\log f(z_i\mid\bm{z};\bm{\xi})\frac{\partial}{\partial\bm{\xi}'}\log f(z_i\mid\bm{z};\bm{\xi}).
\end{align*}
Then, as the prior intensified information criterion which also considers the complexity of the prior distribution, we propose
\begin{align*}
{\rm PIIC2} \equiv -\sum_{i=1}^n\log f(z_i\mid\bm{z};\hat{\bm{\xi}}_{\bm{z}}) + {\rm tr}\{\hat{\bm{I}}_{1,\bm{z},\hat{\bm{\xi}}_{\bm{z}}}(\hat{\bm{\theta}}_{\bm{z},\hat{\bm{\xi}}_{\bm{z}}})^{-1} \hat{\bm{I}}_{2,\bm{z},\hat{\bm{\xi}}_{\bm{z}}}(\hat{\bm{\theta}}_{\bm{z},\hat{\bm{\xi}}_{\bm{z}}}) + \hat{\bm{J}}_{1,\bm{z}}(\hat{\bm{\xi}}_{\bm{z}})^{-1} \hat{\bm{J}}_{2,\bm{z}}(\hat{\bm{\xi}}_{\bm{z}})\}.
\end{align*}

\section{Numerical experiments}
\label{sec4}
We numerically checked the usefulness of the proposed method. In order to understand the characteristics of the PIIC in comparison with those of the WAIC, and in view of the high computational burden of Bayesian prediction, we decided to examine a simple setting. In each experiment, we numerically evaluated the prediction performances of the models estimated by the WAIC (WAIC1) and the PIIC (PIIC1) with one regularization parameter, as well as the prediction performances of the models estimated by the WAIC (WAIC2) and the PIIC (PIIC2) with three regularization parameters. The prediction performance was evaluated by the Kullback-Leibler divergence between the Bayesian predictive distribution and the true distribution, which is $\E_{\breve{z}}\{-\log f(\breve{z}\mid\bm{z};\hat{\bm{\xi}}_{\bm{z}})\}$ using the notation in Section \ref{sec3}. As a reference, the rate at which WAIC1 is less than, equal to, or greater than PIIC1 is indicated by Rate1, and the rate at which WAIC2 is less than, equal to, or greater than PIIC2 is indicated by Rate2. In comparing WAIC1 and PIIC1, we wanted to examine the effect of intensifying the prior distribution in the derivation of the information criterion, while in comparing WAIC2 and PIIC2, we wanted to examine the effect of also using a penalty for the complexity of the prior distribution.

The first simulation settingconsidered linear regression when a normal distribution is used as the prior distribution. Specifically, the model is as follows ($i\in\{1,2,\ldots,n\}$, $j\in\{1,2,\ldots,p\}$): 
\begin{align}
y_i=\sum_{j=1}^px_{ij}\theta_j+\varepsilon_i, \qquad \varepsilon_i\stackrel{\rm i.i.d.}{\sim}{\rm N}(0,\sigma^2), \qquad \theta_j\sim{\rm N}(0,\zeta_j), \qquad x_{ij}\stackrel{\rm i.i.d.}{\sim}{\rm N}(0,1).
\label{simu1}
\end{align}
For the true value of the regression coefficient vector $(\theta_1,\theta_2,\ldots,\theta_p)$, we set $(\theta^*_1\bm{1}_{p/3}^{\T},\theta^*_2\bm{1}_{p/3}^{\T},\allowbreak\theta^*_3\bm{1}_{p/3}^{\T})^{\T}$, where $\bm{1}_{p/3}$ is a $(p/3)$-dimensional one vector. For the true distribution $f_{\varepsilon}$ of $\varepsilon_i$, we used not only a normal distribution but also a t-distribution. For the regularization parameters, we set $(\zeta_1,\zeta_2,\ldots,\zeta_p)=\xi\bm{1}_p$ when using one parameter, and $(\zeta_1,\zeta_2,\ldots,\zeta_p)=(\xi_1\bm{1}_{p/3}^{\T},\xi_2\bm{1}_{p/3}^{\T},\xi_3\bm{1}_{p/3}^{\T})^{\T}$ when using three parameters. 

Table \ref{tab1} compares the WAIC and PIIC under this setting. As can be seen, there is not enough difference between WAIC1 and PIIC1to determine which is superior or inferior. We suspect that the asymptotics used in the WAIC do not necessarily give a good approximation because of the relatively large parameter dimension $p$ for the data size $n$, but no negative effect was observed in this setting. On the other hand, the table shows that PIIC2 is better than the WAIC2 in all settings. In particular, when the variance of the noise $\varepsilon$ is large, in other words, when the parameter $\bm{\theta}_j^*$ is small compared to the size of the noise, and when the model assumes a normal distribution but ${\rm t}(2)$ is true, which actually deviates significantly from the normal distribution, the PIIC2 is clearly superior.

\begin{table}[t!]
\caption{Comparison of WAIC and PIIC when a normal distribution is used as the prior distribution in linear regression.}
\begin{center}
\begin{small}
\begin{tabular}{cccccccccc}
\toprule
$n$ & $p$ & $(\theta_1^*,\theta_2^*,\theta_3^*)$ & $f_{\varepsilon}$ & WAIC1 & PIIC1 & Rate1 & WAIC2 & PIIC2 & Rate2
\\ \midrule
12 & 6 & (2,2,2) & N(0,0.5) & 0.458 & 0.458 & (46,16,38)
 & 0.386 & 0.382 & (42,3,55) 
\\
12 & 6 & (2,2,2) & N(0,1) & 0.456 & 0.456 & (48,4,48)
 & 0.484 & 0.460 & (31,1,68) 
\\
12 & 6 & (2,2,2) & N(0,2) & 0.457 & 0.455 & (46,3,51)
 & 0.590 & 0.465 & (5,0,95) 
\\
12 & 6 & (2,2,2) & t(5) & 0.818 & 0.814 & (41,8,51)
 & 0.923 & 0.828 & (14,1,85) 
\\
12 & 6 & (2,2,2) & t(2) & 3.571 & 3.558 & (31,20,49)
 & 3.689 & 3.591 & (17,8,75) 
\\
12 & 6 & (3,2,1) & N(0,1) & 0.459 & 0.457 & (41,14,45)
 & 0.477 & 0.463 & (43,1,56) 
\\
12 & 6 & (3,1,$-$1) & N(0,1) & 0.462 & 0.460 & (41,13,46)
 & 0.475 & 0.464 & (43,1,56) 
\\
12 & 9 & (2,2,2) & N(0,1) & 0.515 & 0.513 & (45,9,46)
 & 0.622 & 0.525 & (25,2,73) 
\\
18 & 9 & (2,2,2) & N(0,1) & 0.518 & 0.524 & (59,0,41)
 & 0.546 & 0.524 & (37,0,63) 
\\
18 & 12 & (2,2,2) & N(0,0.5) & 0.505 & 0.507 & (55,5,40)
 & 0.459 & 0.454 & (46,0,54) 
\\
18 & 12 & (2,2,2) & N(0,1) & 0.495 & 0.495 & (50,4,46)
 & 0.535 & 0.497 & (33,0,67) 
\\
18 & 12 & (2,2,2) & N(0,2) & 0.489 & 0.487 & (46,3,51)
 & 0.616 & 0.496 & (11,0,89) 
\\
18 & 12 & (2,2,2) & t(5) & 0.753 & 0.764 & (56,4,40)
 & 0.814 & 0.764 & (31,0,69) 
\\
18 & 12 & (2,2,2) & t(2) & 4.159 & 4.117 & (45,21,34)
 & 4.323 & 4.195 & (23,8,69) 
\\
18 & 12 & (3,2,1) & N(0,1) & 0.509 & 0.505 & (48,7,45)
 & 0.522 & 0.507 & (41,0,59) 
\\
18 & 12 & (3,1,$-$1) & N(0,1) & 0.509 & 0.500 & (24,41,35)
 & 0.524 & 0.501 & (36,4,60) 
\\
18 & 15 & (2,2,2) & N(0,1) & 0.447 & 0.433 & (43,9,48)
 & 0.526 & 0.441 & (20,1,79) 
\\
24 & 15 & (2,2,2) & N(0,1) & 0.454 & 0.464 & (53,1,46)
 & 0.478 & 0.464 & (39,1,60) 
\\
24 & 18 & (2,2,2) & N(0,0.5) & 0.534 & 0.540 & (57,3,40)
 & 0.510 & 0.505 & (47,0,53) 
\\
24 & 18 & (2,2,2) & N(0,1) & 0.519 & 0.531 & (56,2,42)
 & 0.572 & 0.532 & (36,0,64) 
\\
24 & 18 & (2,2,2) & N(0,2) & 0.501 & 0.514 & (52,2,46)
 & 0.639 & 0.515 & (13,0,87) 
\\
24 & 18 & (2,2,2) & t(5) & 0.812 & 0.822 & (51,7,42)
 & 0.878 & 0.827 & (36,2,62) 
\\
24 & 18 & (2,2,2) & t(2) & 3.021 & 3.032 & (46,20,34)
 & 3.160 & 3.045 & (28,4,68) 
\\
24 & 18 & (3,2,1) & N(0,1) & 0.543 & 0.528 & (40,14,46)
 & 0.563 & 0.534 & (35,0,65) 
\\
24 & 18 & (3,1,$-$1) & N(0,1) & 0.559 & 0.543 & (18,41,41)
 & 0.585 & 0.543 & (34,6,60) 
\\ \bottomrule
\end{tabular}
\end{small}
\end{center}
\label{tab1}
\end{table}

The second simulation setting considered linear regression when a Laplace distribution was used as the prior distribution. Specifically, in the model \eqref{simu1}, we used $\theta_j\sim{\rm Laplace}(0,\zeta_j)$ instead of $\theta_j\sim{\rm N}(0,\zeta_j)$. The true value of the regression coefficient vector, the true distribution of the noise, and the placement of the regularization parameters were the same as in the first setting. Table \ref{tab2} comapres the WAIC and PIIC under this setting. This time, PIIC1 outperforms WAIC1 in almost all cases. In particular, the difference is especially clear for when the noise distribution is misspecifiedr. Moreover, although the PIIC2 is not superior to WAIC2 in every case, its superiority is marked especially when the noise distribution is misspecified. Another observation is that when the true value $(\theta_1^*,\theta_2^*,\theta_3^*)$ is $(4,0,-2)$ or $(3,1,-1)$, the PIIC is not clearly superior. For such true values, both the WAIC and PIIC give greatly shrunken estimates. Also, when $n$ and especially $p$ are large, the PIIC  clearly becomes advantageous. This is probably because the asymptotics used in the WAIC are more broken in this case, but many of the applications of Bayesian methods have large data sizes and parameter dimensions, suggesting the usefulness of the PIIC in such situations.

\begin{table}[t!]
\caption{Comparison of WAIC and PIIC when a Laplace distribution is used as the prior distribution in linear regression.}
\begin{center}
\begin{small}
\begin{tabular}{cccccccccc}
\toprule
$n$ & $p$ & $(\theta_1^*,\theta_2^*,\theta_3^*)$ & $f_{\varepsilon}$ & WAIC1 & PIIC1 & Rate1 & WAIC2 & PIIC2 & Rate2
\\ \midrule
18 & 12 & (3,2,1) & N(0,0.5) & 1.715 & 1.636 & (31,15,54)
 & 2.040 & 1.603 & (36,0,64) 
\\
18 & 12 & (3,2,1) & N(0,1) & 1.816 & 1.753 & (28,13,59)
 & 2.175 & 1.600 & (36,0,64) 
\\
18 & 12 & (3,2,1) & N(0,2) & 1.788 & 1.748 & (25,17,58)
 & 2.669 & 1.610 & (14,0,86) 
\\
18 & 12 & (3,2,1) & t(2) & 6.138 & 4.982 & (16,8,76)
 & 6.235 & 4.921 & (16,0,84) 
\\
18 & 12 & (4,2,0) & N(0,1) & 1.631 & 1.613 & (22,26,52)
 & 1.532 & 1.447 & (47,0,53) 
\\
18 & 12 & (4,0,$-$2) & N(0,1) & 1.767 & 1.810 & (26,56,18)
 & 1.536 & 1.764 & (64,0,36) 
\\
18 & 12 & (3,1,$-$1) & N(0,1) & 2.108 & 2.174 & (20,61,19)
 & 2.137 & 2.194 & (54,0,46) 
\\
18 & 15 & (3,2,1) & N(0,1) & 2.742 & 2.694 & (24,20,46)
 & 4.029 & 2.630 & (35,0,65) 
\\
24 & 12 & (3,2,1) & N(0,1) & 1.365 & 1.338 & (30,20,50)
 & 1.513 & 1.243 & (29,0,71) 
\\
24 & 15 & (3,2,1) & N(0,0.5) & 1.464 & 1.383 & (25,9,66)
 & 1.753 & 1.359 & (21,0,79) 
\\
24 & 15 & (3,2,1) & N(0,1) & 1.637 & 1.588 & (21,15,64)
 & 1.880 & 1.378 & (25,0,75) 
\\
24 & 15 & (3,2,1) & N(0,2) & 1.503 & 1.482 & (18,25,57)
 & 2.342 & 1.332 & (5,0,95) 
\\
24 & 15 & (3,2,1) & t(2) & 5.920 & 5.096 & (12,15,73)
 & 6.422 & 5.047 & (20,0,80) 
\\
24 & 15 & (4,2,0) & N(0,1) & 1.436 & 1.317 & (15,24,61)
 & 1.554 & 1.226 & (33,0,67) 
\\
24 & 15 & (4,0,$-$2) & N(0,1) & 1.699 & 1.584 & (17,62,21)
 & 1.590 & 1.504 & (58,0,42) 
\\
24 & 15 & (3,1,$-$1) & N(0,1) & 2.135 & 2.103 & (29,54,17)
 & 1.984 & 1.932 & (57,0,43) 
\\
24 & 18 & (3,2,1) & N(0,1) & 2.222 & 2.016 & (20,23,57)
 & 2.896 & 1.924 & (26,0,74) 
\\
30 & 15 & (3,2,1) & N(0,1) & 1.201 & 1.099 & (23,16,61)
 & 1.344 & 1.048 & (26,0,74) 
\\
30 & 18 & (3,2,1) & N(0,0.5) & 1.273 & 1.255 & (38,19,43)
 & 1.653 & 1.258 & (30,0,70) 
\\
30 & 18 & (3,2,1) & N(0,1) & 1.327 & 1.288 & (30,28,42)
 & 1.631 & 1.217 & (30,0,70) 
\\
30 & 18 & (3,2,1) & N(0,2) & 1.411 & 1.362 & (22,35,43)
 & 1.822 & 1.332 & (19,0,81) 
\\
30 & 18 & (3,2,1) & t(2) & 4.645 & 3.942 & (15,6,79)
 & 5.514 & 4.030 & (6,0,94) 
\\
30 & 18 & (4,2,0) & N(0,1) & 1.276 & 1.271 & (19,41,40)
 & 1.397 & 1.119 & (31,0,69) 
\\
30 & 18 & (4,0,$-$2) & N(0,1) & 1.480 & 1.373 & (18,49,33)
 & 1.460 & 1.323 & (47,0,53) 
\\
30 & 18 & (3,1,$-$1) & N(0,1) & 1.915 & 1.822 & (18,34,48)
 & 1.843 & 1.754 & (48,0,52) 
\\ \bottomrule
\end{tabular}
\end{small}
\end{center}
\label{tab2}
\end{table}

The third simulation setting considered logistic regression when a Laplace distribution is used as the prior distribution. Specifically, the model is as follows ($i\in\{1,2,\ldots,n\}$, $j\in\{1,2,\ldots,p\}$):
\begin{align*}
y_i\stackrel{\rm indep.}{\sim}{\rm Binomial}(m,p_i), \quad p_i=\frac{\exp(\sum_{j=1}^px_{ij}\theta_j)}{1+\exp(\sum_{j=1}^px_{ij}\theta_j)}, \quad \theta_j\sim{\rm N}(0,\zeta_j), \quad x_{ij}\stackrel{\rm i.i.d.}{\sim}{\rm N}(0,1).
\end{align*}
For the true value of the regression coefficient vector $(\theta_1,\theta_2,\ldots,\theta_p)$, we set $(\theta^*_1\bm{1}_{p/3}^{\T},\theta^*_2\bm{1}_{p/3}^{\T},\allowbreak\theta^*_3\bm {1}_{p/3}^{\T})^{\T}$. As the true functional form $f_{p}$ of $p_i$, we also considered the probit $\Phi^{-1}\{\exp(\sum_{j=1}^p\allowbreak x_{ij}\theta_j)\}$, where $\Phi(\cdot)$ is the distribution function of the standard normal distribution. The placement of the regularization parameters was the same as in the first and second settings. Table \ref{tab3} compares the WAIC and PIIC under this setting. 

\begin{table}[t!]
\caption{Comparison of WAIC and PIIC when a Laplace distribution is used as the prior distribution in logistic regression.}
\begin{center}
\begin{small}
\begin{tabular}{ccccccccccc}
\toprule
$n$ & $p$ & $(\theta_1^*,\theta_2^*,\theta_3^*)$ & $m$ & $f_p$ & WAIC1 & PIIC1 & Rate1 & WAIC2 & PIIC2 & Rate2
\\ \midrule
20 & 6 & (3,2,1) & 5 & logit & 1.540 & 1.455 & (21,36,38)
 & 1.529 & 1.459 & (33,0,62) 
\\
20 & 6 & (3,2,1) & 10 & logit & 2.619 & 2.484 & (22,10,68)
 & 3.030 & 2.509 & (26,0,74) 
\\
20 & 6 & (3,2,1) & 15 & logit & 3.893 & 3.745 & (21,4,73)
 & 3.976 & 3.754 & (22,0,76) 
\\
20 & 6 & (4,2,0) & 10 & logit & 2.138 & 2.092 & (30,35,34)
 & 2.121 & 2.093 & (44,0,55) 
\\
20 & 6 & (4,0,$-$2) & 10 & logit & 2.409 & 2.290 & (18,42,39)
 & 2.320 & 2.293 & (46,0,53) 
\\
20 & 6 & (3,2,1) & 10 & prob & 4.624 & 4.068 & (5,0,95)
 & 4.548 & 4.063 & (6,0,94) 
\\
20 & 9 & (3,2,1) & 10 & logit & 2.553 & 2.481 & (17,41,36)
 & 2.480 & 2.440 & (36,0,58) 
\\
20 & 9 & (3,2,1) & 10 & prob & 4.820 & 4.171 & (8,1,88)
 & 4.802 & 4.167 & (6,0,91) 
\\
30 & 6 & (3,2,1) & 10 & logit & 2.432 & 2.380 & (23,25,52)
 & 2.410 & 2.388 & (42,0,58) 
\\
30 & 6 & (3,2,1) & 10 & prob & 4.553 & 4.119 & (7,0,93)
 & 4.497 & 4.131 & (12,0,88) 
\\
30 & 9 & (3,2,1) & 5 & logit & 1.410 & 1.292 & (16,29,45)
 & 1.309 & 1.273 & (39,0,51) 
\\
30 & 9 & (3,2,1) & 10 & logit & 2.124 & 2.097 & (25,27,44)
 & 2.116 & 2.096 & (44,0,52) 
\\
30 & 9 & (3,2,1) & 15 & logit & 3.264 & 3.087 & (9,5,85)
 & 3.790 & 3.100 & (11,0,88) 
\\
30 & 9 & (4,2,0) & 10 & logit & 1.924 & 1.833 & (12,29,53)
 & 1.902 & 1.835 & (32,0,62) 
\\
30 & 9 & (4,0,$-$2) & 10 & logit & 2.096 & 2.010 & (15,33,51)
 & 2.037 & 1.998 & (48,0,51) 
\\
30 & 9 & (3,2,1) & 10 & prob & 4.725 & 4.160 & (6,0,94)
 & 4.768 & 4.166 & (3,0,97) 
\\
30 & 12 & (3,2,1) & 10 & logit & 2.082 & 1.976 & (13,36,43)
 & 2.018 & 1.963 & (34,0,58) 
\\
30 & 12 & (3,2,1) & 10 & prob & 4.534 & 3.944 & (5,0,95)
 & 4.594 & 3.934 & (5,0,95) 
\\
40 & 9 & (3,2,1) & 10 & logit & 2.163 & 2.038 & (17,1,82)
 & 2.304 & 2.072 & (24,0,76) 
\\
40 & 9 & (3,2,1) & 10 & prob & 4.542 & 4.061 & (7,0,93)
 & 4.533 & 4.067 & (6,0,94) 
\\
40 & 12 & (3,2,1) & 5 & logit & 1.246 & 1.175 & (15,26,39)
 & 1.206 & 1.174 & (40,0,40) 
\\
40 & 12 & (3,2,1) & 10 & logit & 2.018 & 1.950 & (29,23,45)
 & 1.995 & 1.956 & (40,0,57) 
\\
40 & 12 & (3,2,1) & 15 & logit & 2.831 & 2.750 & (14,26,59)
 & 2.799 & 2.742 & (31,0,68) 
\\
40 & 12 & (4,2,0) & 10 & logit & 1.692 & 1.625 & (22,32,43)
 & 1.636 & 1.601 & (42,0,55) 
\\
40 & 12 & (4,0,$-$2) & 10 & logit & 1.752 & 1.676 & (11,30,51)
 & 1.722 & 1.658 & (29,0,63) 
\\
40 & 12 & (3,2,1) & 10 & prob & 4.594 & 3.985 & (2,0,98)
 & 4.568 & 4.007 & (4,0,96) 
\\ \bottomrule
\end{tabular}
\end{small}
\end{center}
\label{tab3}
\end{table}

In this table, PIIC1 outperforms WAIC1 in all cases. In particular, when the true link function is misspecified as the probit, the difference becomes more pronounced. Moreover, PIIC2 outperforms WAIC2 in all cases, and the difference is most apparent in the case of misspecification. On the other hand, unlike in Tables \ref{tab1} and \ref{tab2}, the difference between PIIC2 and WAIC2 is not much larger than the difference between PIIC1 and WAIC1. This means that the effect of considering an asymptotic theory that intensifies the prior distribution is larger in this setting than the effect of considering the complexity of the prior distribution. In any case, in these three settings, where the parameter dimension is relatively large in proportion to the sample size, the PIIC tends to have a smaller prediction squared error than that of the WAIC, and in some cases, the PIIC clearly outperforms the WAIC.

\section{Real data analysis}
\label{sec5}
We applied the proposed method to real data to check how much difference it makes from the existing methods. The diabetes data presented here consist of $p=10$ baseline explanatory variables for $n=442$ patients, including age, sex, body mass index, average blood pressure, and six blood serum measurements, as well as an objective variable indicating the progression of the disease one year after the baseline. \cite{EfrHJT04} proposed the least-angle regression (LARS) method to accurately predict future patient status at the baseline and to identify which explanatory variables are important factors in disease progression. They constructed a model to predict the objective variable $y$ from the explanatory variables $x_1,x_2,\ldots,x_{10}$. Considering the same purpose, we supposed variable selection by sparse estimation and performed a linear regression analysis using a Laplace distribution as the prior distribution. These data have a sufficiently large $n$ compared with $p$, and in that sense are not the subject of this paper. Therefore, we divided the data into 13 sets of size $n=34$ and made inferences on each of them.

The model and prior distributions were set up in the same way as in Section \ref{sec4}, and considering the type of explanatory variables, two regularization parameters were used: one for the regression coefficients of the six serum measurements and another for the regression coefficients of the other four. That is, $\theta_1,\theta_2,\theta_3,\theta_4\sim{\rm Laplace}(0,\xi_1)$ and $\theta_5,\theta_6,\ldots,\theta_{10}\sim{\rm Laplace}(0,\xi_2)$. Table \ref{tab4} compares the Bayesian estimator of $(\theta_1,\theta_2,\ldots,\theta_{10})$ from WAIC2 with that from PIIC2 in this setting. Since there are 13 sets of data, there are 13 sets of 10-dimensional estimators for each criterion. Since there is no need to use WAIC1 and PIIC1 if the method has more than one regularization parameter, we have not included their comparison as in Section \ref{sec4}.

\begin{table}[t!]
\caption{Comparison of estimators when a Laplace distribution is used as the prior distribution in linear regression analysis for diabetes data.}
\begin{center}
\begin{small}
{\tabcolsep=1.8mm
\begin{tabular}{crrrrrrrrrr}
\toprule
 & \multicolumn{1}{c}{age} & \multicolumn{1}{c}{sex} & \multicolumn{1}{c}{bmi} & \multicolumn{1}{c}{map} & \multicolumn{1}{c}{tc} & \multicolumn{1}{c}{ldl} & \multicolumn{1}{c}{hdl} & \multicolumn{1}{c}{tch} & \multicolumn{1}{c}{ltg} & \multicolumn{1}{c}{glu}
\\ \midrule
WAIC2 & 0.00 & 0.00 & 0.00 & 0.43 & $-$44.17 & 17.66 & 20.85 & 33.89 & 47.88 & $-$7.65 
\\
PIIC2 & 0.00 & 0.00 & 8.44 & 0.00 & 0.00 & 0.00 & $-$3.18 & 0.00 & 24.31 & 0.00 
\\
WAIC2 & 6.06 & $-$9.91 & 13.85 & 6.87 & 0.00 & $-$1.11 & 2.11 & 0.00 & 19.48 & $-$4.58 
\\
PIIC2 & 0.00 & 0.00 & 7.00 & 3.16 & 0.00 & 0.00 & 0.00 & 0.00 & 13.14 & 0.00 
\\
WAIC2 & 0.00 & $-$16.30 & 5.35 & 0.00 & 73.40 & $-$48.12 & $-$59.28 & $-$19.91 & $-$1.43 & 5.93 
\\
PIIC2 & 0.00 & $-$8.92 & 6.53 & 0.00 & 0.00 & $-$2.29 & $-$15.41 & 0.00 & 11.62 & 0.56 
\\
WAIC2 & $-$7.33 & $-$5.04 & 2.67 & 9.15 & 0.00 & $-$1.54 & $-$5.81 & $-$7.05 & 23.78 & 10.63 
\\
PIIC2 & $-$8.90 & $-$9.70 & 8.38 & 31.12 & 0.00 & 0.00 & 0.00 & 0.00 & 1.54 & 0.00 
\\
WAIC2 & 7.84 & $-$6.67 & 30.60 & 8.76 & 0.00 & 0.00 & 0.00 & 9.96 & 5.10 & 1.07 
\\
PIIC2 & 1.75 & 0.00 & 25.07 & 4.57 & 0.00 & 0.00 & 0.00 & 6.22 & 9.20 & 4.40 
\\
WAIC2 & 3.64 & $-$0.59 & 16.17 & 0.91 & $-$0.40 & $-$5.99 & $-$2.32 & 0.00 & 10.86 & 5.51 
\\
PIIC2 & 0.00 & 0.00 & 14.04 & 0.00 & 0.00 & 0.00 & 0.00 & 0.00 & 9.50 & 4.82 
\\
WAIC2 & 0.00 & 1.53 & 3.34 & 0.00 & 0.00 & 0.33 & $-$8.10 & 1.39 & 15.93 & 9.43 
\\
PIIC2 & 0.00 & 0.47 & 0.00 & 0.00 & 0.00 & 0.00 & $-$7.69 & 3.39 & 18.31 & 13.16 
\\
WAIC2 & 0.00 & 0.00 & 33.50 & 7.03 & 0.00 & $-$8.73 & $-$5.02 & 0.00 & 22.16 & 3.98 
\\
PIIC2 & 0.00 & 0.00 & 35.40 & 9.90 & 0.00 & 0.00 & $-$1.55 & 0.00 & 17.20 & 0.00 
\\
WAIC2 & 1.20 & 0.00 & 0.40 & 0.00 & 1.71 & 27.08 & $-$12.72 & $-$20.40 & 28.97 & $-$2.60 
\\
PIIC2 & 16.00 & $-$9.38 & 19.27 & 10.50 & 0.00 & 0.00 & 0.00 & 8.17 & 12.04 & 0.00 
\\
WAIC2 & 2.88 & 0.00 & 11.06 & 4.03 & 0.00 & 0.00 & 0.00 & 9.97 & 0.00 & 0.00 
\\
PIIC2 & 0.00 & 0.00 & 3.80 & 0.00 & 0.00 & 0.00 & 0.00 & 18.19 & 0.00 & 0.00 
\\
WAIC2 & 1.51 & 0.00 & 12.88 & 0.00 & $-$40.47 & 48.67 & $-$12.47 & $-$42.42 & 41.70 & 16.40 
\\
PIIC2 & 0.05 & 0.00 & 19.26 & 5.58 & 0.00 & 0.00 & 0.00 & 0.00 & 0.00 & 14.74 
\\
WAIC2 & 0.00 & 0.00 & 12.50 & 5.34 & 0.00 & 0.00 & $-$12.18 & 0.00 & 17.47 & $-$0.70 
\\
PIIC2 & 0.00 & 0.00 & 14.31 & 6.14 & 0.00 & 0.00 & $-$10.82 & 0.00 & 15.40 & 0.00 
\\
WAIC2 & $-$4.20 & $-$7.32 & 17.82 & 19.93 & 0.00 & 0.00 & $-$1.07 & 0.00 & 16.67 & 0.00 
\\
PIIC2 & 0.00 & 0.00 & 14.60 & 10.87 & 0.00 & 0.00 & 0.00 & 0.00 & 16.97 & 0.00 
\\ \bottomrule
\end{tabular}
}
\end{small}
\end{center}
\label{tab4}
\end{table}

The fact that the estimates are zero means that the corresponding explanatory variables were not selected, so we can see that the criteria give very different results even for variable selection. In addition, although the 9-th and 10-th lines select almost the same variables, the estimates are still quite different. Since this is an analysis of real data, the true value is of course not known, and it is not possible to argue which criterion is more appropriate, but it does indicate that improving the criterion in this type of analysis can make a big difference.

\section{Extension}
\label{sec6}
Although we have dealt with the basic and general settings, we should be aware that information criteria have developed in various directions, so an important task will be to develop the PIIC in the same way. One setting in which the classical criteria have been reevaluated and found to have significant differences is that of causal inference (\citealt{BabKN17}). In this section, we show how the PIIC can be modified when using the inverse probability weighted estimation (\citealt{RobRZ94}) in propensity score analysis and suggest that such modifications may be possible in various other settings.

Let us consider a marginal structural model ($i\in\{1,2,\ldots,n\}$), which is a standard one for causal inference:
\begin{align*}
y_i = \sum_{h=1}^H t_i^{(h)} y_i^{(h)}, \qquad y_i^{(h)}\stackrel{\rm i.i.d.}{\sim} f^{(h)}(\cdot \mid \bm{\theta}), \qquad \bm{\theta}\sim\pi_n(\cdot;\bm{\xi}).
\end{align*}
Here, $t_i^{(h)}\; (\in\{0,1\})$ is the assignment variable that is set to 1 when the $h$-th treatment is assigned ($\sum_{h=1}^H t_i^{(h)}=1$), $y_i^{(h)}\; (\in\mathbb{R})$ is the outcome variable when the $h$-th treatment is assigned, $f^{(h)}$ is the distribution of $y_i^{(h)}$ ($h\in\{1,2,\ldots,H\}$). In many causal inferences, the setup is $f^{(h)}(\cdot \mid \bm{\theta})=f(\cdot \mid \bm{\theta}^{(h)})$, but for our purposes, it is simpler to summarize $\bm{\theta}^{(h)}$, so we will write it that way. Note that $y_i$ on the left-hand side is the observed outcome variable. In this model, $y_i^{(h)}$ with $t_i^{(h)}=0$, i.e., the $H-1$ latent outcome variables, are considered to be missing. Moreover, since $\E(y_i^{(h)})\neq \E(y_i^{(h)}\mid t_i^{(h)}=1)$ in general, simply constructing the predictive distribution from the observed values alone will result in a bias. Here, we assume that the confounding variables $\bm{x}_i\ (\in\mathbb{R}^s)$ between $y_i^{(h)}$ and $t_i^{(h)}$ are observed so that this bias can be removed. In addition, we assume an ignorable treatment assignment condition, to allow the removal of this bias, 
\begin{align*}
y_i^{(h)} \indep t_i^{(h)} \mid \bm{x}_i \qquad (h\in\{1,2,\ldots,H\}).
\end{align*}
Moreover, we assume positivity condition , $\P(t_i^{(h)}=1\mid\bm{x}_i)>0$.

In the inverse probability weighted estimation, defining the propensity score as $e^{(h)}(\bm{x}_i)\equiv\P(t_i^{(h)}=1\mid\bm{x}_i)$, the weighted log-likelihood $\sum_{h=1}^H\sum_{i=1}^n\{t_i^{(h)}/e^{(h)}(\bm{x}_i)\}\log f^{(h)}(y_i^{(h)}\mid\bm{\theta})=\log\{\prod_{h=1}^H\prod_{i=1}^nf^{(h)}(y_i^{(h)}\mid\bm{\theta})^{t_i^{(h)}/e^{(h)}(\bm{x}_i)}\}$ is maximized. Consequently, we can construct the predictive distribution by using the inverse probability weighting, as
\begin{align}
& f_{\rm IP}(\breve{y}\mid\bm{z},\tilde{\bm{x}};\bm{\xi})
\notag \\
& \equiv \frac{\int \prod_{h=1}^Hf^{(h)}(\breve{y}^{(h)}\mid\bm{\theta})^{\breve{t}^{(h)}/e^{(h)}(\breve{\bm{x}})} \prod_{h=1}^H\prod_{i=1}^nf^{(h)}(y_i^{(h)}\mid\bm{\theta})^{t_i^{(h)}/e^{(h)}(\bm{x}_i)} \pi(\bm{\theta};\bm{\xi})^{n/n_0} {\rm d}\bm{\theta}}{\int \prod_{h=1}^H\prod_{i=1}^nf^{(h)}(y_i^{(h)}\mid\bm{\theta})^{t_i^{(h)}/e^{(h)}(\bm{x}_i)} \pi(\bm{\theta};\bm{\xi})^{n/n_0} {\rm d}\bm{\theta}}.
\label{ipwpd}
\end{align}
Here, $\bm{z}=(y_1,y_2,\ldots,y_n,\bm{x}_1,\bm{x}_2,\ldots,\bm{x}_n)$, and $\breve{\bm{z}}=(\breve{y},\breve{\bm{x}})$ is prepared to handle the data of $\bm{z}_i=(y_i,\bm{x}_i)$ and their future data $\tilde{\bm{z}}_i=(\tilde{y}_i,\tilde{\bm{x}}_i)$ in a unified manner t. Note that $y_i^{(h)}$ and $t_i^{(h)}$ appear on the right-hand side of \eqref{ipwpd}, for example, but the left-hand side is denoted as such because it depends only on $y_i$ as a result. The logarithm of this predictive distribution is the one corresponding to \eqref{logpd}. The risk is $-\sum_{i=1}^n\E_{\tilde{\bm{z}}_i}\{\log f_{\rm IP}(\tilde{y}_i\mid\bm{z},\tilde{\bm{x}}_i;\bm{\xi})\}$ based on this predictive distribution. For this risk, let $-\sum_{i=1}^n\log f_{\rm IP}(y_i\mid\bm{z},\bm{x}_i;\bm{\xi})$ be the initial evaluation, and use the asymptotic evaluation of the expectation of
\begin{align}
& \sum_{i=1}^n\log f_{\rm IP}(y_i\mid\bm{z},\bm{x}_i;\bm{\xi})
-\sum_{i=1}^n\sum_{h=1}^H\frac{t_i^{(h)}}{e^{(h)}(\bm{x}_i)}\log f^{(h)}(y_i^{(h)}\mid\bm{\theta}^*_{\bm{\xi}})
\notag \\
& -\sum_{i=1}^n\log f_{\rm IP}(\tilde{y}_i\mid\bm{z},\tilde{\bm{x}}_i;\bm{\xi})
+\sum_{i=1}^n\sum_{h=1}^H\frac{\tilde{t}_i^{(h)}}{e^{(h)}(\tilde{\bm{x}}_i)}\log f^{(h)}(\tilde{y}_i^{(h)}\mid\bm{\theta}^*_{\bm{\xi}})
\label{orib}
\end{align}
for the correction. The first and third terms $\log f_{\rm IP}(\breve{y}\mid\bm{z},\breve{\bm{x}};\bm{\xi})$ are the integrals in \eqref{ipwpd}, and if we apply a Laplace approximation to these integrals, we get
\begin{align*}
\sum_{h=1}^H\sum_{i=1}^n\log\frac{g_{\rm IP}^{(h)}(\bm{z}_i,\hat{\bm{\theta}}_{\breve{\bm{z}},\bm{z},\bm{\xi}};\bm{\xi})}{g_{\rm IP}^{(h)}(\bm{z}_i,\hat{\bm{\theta}}_{\bm{z},\bm{\xi}};\bm{\xi})}-\frac{1}{2}\log\frac{|\hat{\bm I}_{1,\breve{\bm{y}},\bm{y},\bm{\xi}}(\hat{\bm{\theta}}_{\breve{\bm{z}},\bm{z},\bm{\xi}})|}{|\hat{\bm I}_{1,\bm{y},\bm{\xi}}(\hat{\bm{\theta}}_{\bm{z},\bm{\xi}})|}+\sum_{h=1}^H\frac{\breve{t}^{(h)}}{e^{(h)}(\breve{\bm{x}})}\log f^{(h)}(\breve{y}^{(h)}\mid\hat{\bm{\theta}}_{\breve{\bm{z}},\bm{z},\bm{\xi}})
\end{align*}
plus $\OP(n^{-2})$. Here, $\breve{\bm{y}}=(\breve{y}^{(1)},\breve{y}^{(2)},\ldots,\breve{y}^{(H)})$, $\bm{y}_i=(y_i^{(1)},y_i^{(2)},\allowbreak\ldots,y_i^{(H)})$ and $\bm{y}=(\bm{y}_1,\bm{y}_2,\allowbreak\ldots,\bm{y}_n)$, and 
\begin{align}
& g_{\rm IP}^{(h)}(\bm{z}_i,\bm{\theta};\bm{\xi})\equiv f^{(h)}(y_i^{(h)}\mid\bm{\theta})^{t_i^{(h)}/e^{(h)}(\bm{x}_i)} \pi(\bm{\theta};\bm{\xi})^{1/(n_0H)}, 
\notag \\
& \hat{\bm{\theta}}_{\breve{\bm{z}},\bm{z},\bm{\xi}}\equiv\argmax_{\bm{\theta}}\bigg[\sum_{h=1}^H\bigg\{\sum_{i=1}^n\log g_{\rm IP}^{(h)}(\bm{z}_i,\bm{\theta};\bm{\xi})+\frac{\breve{t}^{(h)}}{e^{(h)}(\breve{\bm{x}})}\log f^{(h)}(\breve{y}^{(h)}\mid\bm{\theta})\bigg\}\bigg],
\notag \\
& \hat{\bm{\theta}}_{\bm{z},\bm{\xi}}\equiv\argmax_{\bm{\theta}}\bigg\{\sum_{h=1}^H\sum_{i=1}^n\log g_{\rm IP}^{(h)}(\bm{z}_i,\bm{\theta};\bm{\xi})\bigg\}
\notag \\
& \hat{\bm{I}}_{1,\breve{\bm{z}},\bm{z},\bm{\xi}}(\bm{\theta})\equiv-\frac{1}{n}\frac{\partial^2}{\partial{\bm{\theta}}\partial{\bm{\theta}}'}\sum_{h=1}^H\bigg\{\sum_{i=1}^n\log g_{\rm IP}^{(h)}(\bm{z}_i,\bm{\theta};\bm{\xi})+\frac{\breve{t}^{(h)}}{e^{(h)}(\breve{\bm{x}})}\log f^{(h)}(\breve{y}^{(h)}\mid\bm{\theta})\bigg\}, 
\notag \\
& \hat{\bm{I}}_{1,\bm{z},\bm{\xi}}(\bm{\theta})\equiv-\frac{1}{n}\frac{\partial^2}{\partial{\bm{\theta}}\partial{\bm{\theta}}'}\sum_{h=1}^H\sum_{i=1}^n\log g_{\rm IP}^{(h)}(\bm{z}_i,\bm{\theta};\bm{\xi}). 
\label{hatI1}
\end{align}
From the above and $\hat{\bm{\theta}}_{\breve{\bm{z}},\bm{z},\bm{\xi}}-\hat{\bm{\theta}}_{\bm{z},\bm{\xi}}=\OP(n^{-1})$, it is trivially true that
\begin{align*}
\log f_{\rm IP}(\breve{y}\mid\bm{z},\breve{\bm{x}};\bm{\xi}) = \sum_{h=1}^H\frac{\breve{t}^{(h)}}{e^{(h)}(\breve{\bm{x}})}\log f^{(h)}(\breve{y}^{(h)}\mid\hat{\bm{\theta}}_{\bm{z},\bm{\xi}})+\OP(n^{-1}). 
\end{align*}
Then, as in Section \ref{sec3_1}, the term on the right-hand side comes into play as a bias, and \eqref{orib} is evaluated as
\begin{align}
& \sum_{i=1}^n\sum_{h=1}^H\frac{t_i^{(h)}}{e^{(h)}(\bm{x}_i)}\log f^{(h)}(y_i^{(h)}\mid\hat{\bm{\theta}}_{\bm{z},\bm{\xi}})
-\sum_{i=1}^n\sum_{h=1}^H\frac{t_i^{(h)}}{e^{(h)}(\bm{x}_i)}\log f^{(h)}(y_i^{(h)}\mid\bm{\theta}^*_{\bm{\xi}})
\notag \\
& -\sum_{i=1}^n\sum_{h=1}^H\frac{\tilde{t}_i^{(h)}}{e^{(h)}(\tilde{\bm{x}}_i)}\log f^{(h)}(\tilde{y}_i^{(h)}\mid\hat{\bm{\theta}}_{\bm{z},\bm{\xi}})
+\sum_{i=1}^n\sum_{h=1}^H\frac{\tilde{t}_i^{(h)}}{e^{(h)}(\tilde{\bm{x}}_i)}\log f^{(h)}(\tilde{y}_i^{(h)}\mid\bm{\theta}^*_{\bm{\xi}})+\oP(1).
\label{orib2}
\end{align}
If we apply the conventional asymptotic theory of M-estimation to $\hat{\bm{\theta}}_{\bm{z},\bm{\xi}}$, we obtain $n^{1/2}(\hat{\bm{\theta}}_{\bm{z},\bm{\xi}}-\bm{\theta}^*_{\bm{\xi}})=\hat{\bm{I}}_{1,\bm{z},\bm{\xi}}(\bm{\theta}^*_{\bm{\xi}})^{-1}\allowbreak n^{-1/2}\sum_{i=1}^n\sum_{h=1}^H(\partial/\partial{\bm{\theta}})\log g_{\rm IP}^{(h)}(\bm{z}_i,\bm{\theta}^*_{\bm{\xi}};\bm{\xi})$, so by using this in the Taylor expansion of \eqref{orib2},we get
\begin{align*}
& \frac{1}{n}\sum_{i,j=1}^n\sum_{h,k=1}^H\frac{t_i^{(h)}}{e^{(h)}(\bm{x}_i)}\frac{\partial}{\partial\bm{\theta}'}\log f^{(h)}(y_i^{(h)}\mid\bm{\theta}^*_{\bm{\xi}}) \hat{\bm{I}}_{1,\bm{z},\bm{\xi}}(\bm{\theta}^*_{\bm{\xi}})^{-1} \frac{\partial}{\partial\bm{\theta}}\log g_{\rm IP}^{(k)}(\bm{z}_j,\bm{\theta}^*_{\bm{\xi}};\bm{\xi})
\\
& - \frac{1}{n}\sum_{i,j=1}^n\sum_{h,k=1}^H\frac{\tilde{t}_i^{(h)}}{e^{(h)}(\tilde{\bm{x}}_i)}\frac{\partial}{\partial\bm{\theta}'}\log f^{(h)}(\tilde{y}_i^{(h)}\mid\bm{\theta}^*_{\bm{\xi}}) \hat{\bm{I}}_{1,\bm{z},\bm{\xi}}(\bm{\theta}^*_{\bm{\xi}})^{-1} \frac{\partial}{\partial\bm{\theta}}\log g_{\rm IP}^{(k)}(\bm{z}_j,\bm{\theta}^*_{\bm{\xi}};\bm{\xi})
+ \oP(1).
\end{align*}
Now, letting $\bm{u}_1$, $\bm{u}_2$ and $\bm{u}_3$ be independent random vectors distributed according to ${\rm N}(\bm{0},\allowbreak\bm{I}_{2,\bm{\xi}}(\bm{\theta}^*_{\bm{\xi}}))$, we obtain
\begin{align}
b^{\rm limit} = {\rm tr}\big\{\bm{I}_{1,\bm{\xi}}\big(\bm{\theta}^*_{\bm{\xi}}\big)^{-1}\bm{I}_{2,\bm{\xi}}\big(\bm{\theta}^*_{\bm{\xi}}\big)\big\} + \bm{u}_1'\bm{I}_{1,\bm{\xi}}\big(\bm{\theta}^*_{\bm{\xi}}\big)^{-1}\bm{u}_3 - \bm{u}_2'\bm{I}_{1,\bm{\xi}}\big(\bm{\theta}^*_{\bm{\xi}}\big)^{-1}\bm{u}_3
\label{blim2}
\end{align}
as the weak limit of \eqref{orib}, where
\begin{align}
& \bm{I}_{1,\bm{\xi}}(\bm{\theta}) \equiv \E_{\breve{\bm{y}}}\bigg\{-\sum_{h=1}^H\frac{\partial^2}{\partial\bm{\theta}\partial\bm{\theta}'}\log g^{(h)}(\breve{y}^{(h)},\bm{\theta};\bm{\xi})\bigg\},
\notag \\
& \bm{I}_{2,\bm{\xi}}(\bm{\theta}) \equiv \E_{\breve{\bm{z}}}\bigg\{\sum_{h=1}^H\frac{1}{e^{(h)}(\breve{\bm{x}})}\frac{\partial}{\partial\bm{\theta}}\log g^{(h)}(\breve{y}^{(h)},\bm{\theta};\bm{\xi})\frac{\partial}{\partial\bm{\theta}'}\log g^{(h)}(\breve{y}^{(h)},\bm{\theta};\bm{\xi})\bigg\}. 
\label{Idef2}
\end{align}
Taking the expectation of \eqref{blim2}, the second and third terms on the right side become zero, and we obtain Theorem \eqref{th1} using \eqref{Idef2} instead of \eqref{Idef}. By using \eqref{ipwpd} and \eqref{hatI1} and preparing
\begin{align*}
\hat{\bm{I}}_{2,\bm{z},\bm{\xi}}(\bm{\theta}) \equiv 
\frac{1}{n}\sum_{h=1}^H\sum_{i=1}^n\frac{t_i^{(h)}}{e^{(h)}(\bm{x}_i)^2}\frac{\partial}{\partial\bm{\theta}}\log g_{\rm IP}^{(h)}(\bm{z}_i,\bm{\theta};\bm{\xi})\frac{\partial}{\partial\bm{\theta}'}\log g_{\rm IP}^{(h)}(\bm{z}_i,\bm{\theta};\bm{\xi}),
\end{align*}
we propose
\begin{align*}
{\rm PIIC} \equiv -\sum_{i=1}^n\log f_{\rm IP}(y_i\mid\bm{z},\bm{x}_i;\bm{\xi}) + {\rm tr}\big\{\hat{\bm{I}}_{1,\bm{z},\bm{\xi}}\big(\hat{\bm{\theta}}_{\bm{z},\bm{\xi}}\big)^{-1} \hat{\bm{I}}_{2,\bm{z},\bm{\xi}}\big(\hat{\bm{\theta}}_{\bm{z},\bm{\xi}}\big)\big\}
\end{align*}
as the prior intensified information criterion. Note that, since the propensity score is multiplied by the inverse once in $\hat{\bm{I}}_{1,\bm{z},\bm{\xi}}(\hat{\bm{\theta}}_{\bm{z},\bm{\xi}})$ and twice in $\hat{\bm{I}}_{2,\bm{z},\bm{\xi}}(\hat{\bm{\theta}}_{\bm{z},\bm{\xi}})$, the penalty is several times the one for the PIIC proposed in Section \ref{sec3_1}.

As in Section \ref{sec3_3}, we will evaluate the penalty when the hyper-parameter of the prior distribution is selected as $\hat{\bm{\xi}}_{\bm{z}}$. If we replace $f(z_i\mid\bm{z};\bm{\xi})$ in Section \ref{sec3_3} with $f_{\rm IP}(y_i\mid\bm{z},\bm{x}_i;\bm{\xi})$, the discussion proceeds similarly to Section \ref{sec3_3}, and using
\begin{align*}
& \bm{J}_1(\bm{\xi}) \equiv \E_{\breve{\bm{y}},\bm{z}}\bigg\{-\sum_{h=1}^H\frac{\partial^2}{\partial\bm{\xi}\partial\bm{\xi}'}\log f^{(h)}(\breve{y}^{(h)}\mid\bm{z};\bm{\xi})\bigg\},
\\
& \bm{J}_2(\bm{\xi}) \equiv \E_{\breve{\bm{z}},\bm{z}}\bigg\{\sum_{h=1}^H\frac{1}{e^{(h)}(\breve{\bm{x}})}\frac{\partial}{\partial\bm{\xi}}\log f^{(h)}(\breve{y}^{(h)}\mid\bm{z};\bm{\xi}) \frac{\partial}{\partial\bm{\xi}'}\log f^{(h)}(\breve{y}^{(h)}\mid\bm{z};\bm{\xi})\bigg\}
\end{align*}
instead of \eqref{Jdef} gives us Theorem \ref{th3}. Here, $f^{(h)}(\breve{y}^{(h)}\mid\bm{z};\bm{\xi})$ is the conditional distribution of $\breve{y}^{(h)}$ given $\bm{z}$ as in the notation. Then, preparing 
\begin{align*}
& \hat{\bm{J}}_{1,\bm{z}}(\bm{\xi})\equiv-\frac{1}{n}\sum_{h=1}^H\sum_{i=1}^n\frac{\partial^2}{\partial\bm{\xi}\partial\bm{\xi}'}\frac{t_i^{(h)}}{e^{(h)}(\bm{x}_i)}\log f^{(h)}(y_i^{(h)}\mid\bm{z};\bm{\xi}),
\\
& \hat{\bm{J}}_{2,\bm{z}}(\bm{\xi})\equiv\frac{1}{n}\sum_{h=1}^H\sum_{i=1}^n\frac{t_i^{(h)}}{e^{(h)}(\bm{x}_i)^2}\frac{\partial}{\partial\bm{\xi}}\log f^{(h)}(y_i^{(h)}\mid\bm{z};\bm{\xi}) \frac{\partial}{\partial\bm{\xi}'}\log f^{(h)}(y_i^{(h)}\mid\bm{z};\bm{\xi})
\end{align*}
as consistent estimators for these enables us to propose
\begin{align*}
& {\rm PIIC2} 
\\
& \equiv -\sum_{i=1}^n\log f_{\rm IP}(y_i\mid\bm{z},\bm{x}_i;\hat{\bm{\xi}}_{\bm{z}}) + {\rm tr}\{\hat{\bm{I}}_{1,\bm{z},\hat{\bm{\xi}}_{\bm{z}}}(\hat{\bm{\theta}}_{\bm{z},\hat{\bm{\xi}}_{\bm{z}}})^{-1} \hat{\bm{I}}_{2,\bm{z},\hat{\bm{\xi}}_{\bm{z}}}(\hat{\bm{\theta}}_{\bm{z},\hat{\bm{\xi}}_{\bm{z}}}) + \hat{\bm{J}}_{1,\bm{z}}(\hat{\bm{\xi}}_{\bm{z}})^{-1} \hat{\bm{J}}_{2,\bm{z}}(\hat{\bm{\xi}}_{\bm{z}})\}
\end{align*}
as a prior intensified information criterion that also takes into account the complexity of the prior distribution. Note that, since the propensity score is multiplied by the inverse once in $\hat{\bm{J}}_{1,\bm{z}}(\hat{\bm{\xi}}_{\bm{z}})$ and twice in $\hat{\bm{J}}_{2,\bm{z}}(\hat{\bm{\xi}}_{\bm{z}})$, the penalty is several times the one for the PIIC proposed in Section \ref{sec3_3}.

\section{Conclusion}
\label{sec7}
In this paper, we raised two concerns about the WAIC, which is becoming a standard tool in Bayesian statistics, when it is used for modern complex modeling. Specifically, when the data size $n$ is not large compared with the dimension $p$ of the parameter $\bm{\theta}$, there is a problem in terms of asymptotics wherein the influence of the prior distribution disappears and over-fitting occurs without penalty when the complexity of the prior distribution is also increased. To alleviate these concerns, we derived an information criterion based on asymptotics such that the logarithm of the prior distribution is $\O(n)$ and  derived a penalty term for when the parameters of the prior distribution are determined on the basis of the data. We also showed that when the prior distribution leads to sparse estimation of $\bm{\theta}$, it is sufficient to consider only the terms corresponding to the active set for the penalty of the information criterion, thereby reducing the computational cost. Numerical experiments demonstrated that the PIIC based on the above derivation almost always gives better predictions than the WAIC under the above-mentioned settings and in some cases significantly outperforms it. In addition, we conducted a real data analysis confirming that the actual results of variable selection and estimates differ considerably between the two criteria.

Since the main focus of this paper was to derive a criterion that is appropriate for the above purpose, we have tried to keep the setting as general and as basic as possible; expanding the range of applicable settings will be a major challenge. First of all, in view of numerous applications of Bayesian statistics, it will be essential to remove the assumption of independent samples. This may be an issue not only for the PIIC but also for the WAIC, but it is important to develop a model that includes correlations between samples for use in spatio-temporal statistics. In addition, while we have extended the PIIC to some cases in causal inference to show that it can be developed in the same way as classical information criteria have in various settings, further extensions should be planned. Actually, while we have dealt with the most basic inverse probability weighted estimation, it will also be important to deal with doubly robust estimation (\citealt{BanRob05}), time varying confounding (\citealt{DanCDKS13}) and covariate balancing (\citealt{ImaR14}); they seem to be possible. Furthermore, the challenges discussed in this paper should be overcome in the development of information criteria other than the WAIC, which would be potentially useful in modern Bayesian statistics. They are of particular importance in the field of machine learning, including those proposed in \cite{Wat13}, \cite{VehMTSW16} and \cite{MatUH21}.

The WAIC is also a valid criterion for singular statistical models that cannot be treated with conventional statistical asymptotic theory. Models built using latent variables are often singular and are useful in many applications. However, in singular statistical models, the likelihood ratio statistic does not asymptotically follow a chi-square distribution, and basic information criteria such as AIC do not work well. In this context, \cite{DacG97} and \cite{DacG99} have developed asymptotic distribution theory into a broad framework. Moreover, they have separately treated asymptotic distribution theory for mixture distribution models in a framework that goes even further. On the other hand, \cite{CheCK01} and \cite{CheCK04} have partially removed the singular structure by considering penalized estimation and have derived an asymptotic distribution theory for mixture models with relatively simple results. Other individual theories of likelihood ratio statistics for singular statistical models, such as a neural network model and a factor analysis model, have been addressed in \cite{Fuk03} and \cite{Drt09}. The topics of these papers involve the problem of selecting the number of components, neurons, and factors, but no information criterion such as AIC has been developed. The WAIC avoids some of the singular structures by using Bayesian methods. The PIIC constructed within a Bayesian framework is valid if the prior distribution avoids singular structures, but in cases in which the number of components, neurons, factors, etc. is to be selected, it is necessary to tackle singular structures head-on. Developing an extension to the PIIC that enables selection of the number of components, etc., by incorporating individual theories of likelihood ratio statistics will be an important topic in the future.

\acks{
This work was supported by JSPS Grant-in-Aid for Scientific Research 16K00050.
}

\bibliography{refs}

\begin{thebibliography}{38}
\providecommand{\natexlab}[1]{#1}
\providecommand{\url}[1]{\texttt{#1}}
\expandafter\ifx\csname urlstyle\endcsname\relax
  \providecommand{\doi}[1]{doi: #1}\else
  \providecommand{\doi}{doi: \begingroup \urlstyle{rm}\Url}\fi

\bibitem[Akaike(1973)]{Aka73}
Hirotsugu Akaike.
\newblock Information theory and an extension of the maximum likelihood
  principle.
\newblock In \emph{2nd International Symposium on Information Theory}, pages
  199--213. Akademiai Kiado, 1973.

\bibitem[Akaike(1980)]{Aka80}
Hirotugu Akaike.
\newblock Likelihood and the bayes procedure.
\newblock In \emph{Bayesian Statistics}, pages 143--166. Springer, 1980.

\bibitem[Baba et~al.(2017)Baba, Kanemori, and Ninomiya]{BabKN17}
Takamichi Baba, Takayuki Kanemori, and Yoshiyuki Ninomiya.
\newblock A $c_p$ criterion for semiparametric causal inference.
\newblock \emph{Biometrika}, 104\penalty0 (4):\penalty0 845--861, 2017.

\bibitem[Bang and Robins(2005)]{BanRob05}
Heejung Bang and James~M. Robins.
\newblock Doubly robust estimation in missing data and causal inference models.
\newblock \emph{Biometrics}, 61\penalty0 (4):\penalty0 962--972, 2005.
\newblock ISSN 0006-341X.

\bibitem[Becker et~al.(2017)Becker, Huang, Bieri, Ma, Knowles, Jafar-Nejad,
  Messing, Kim, Soriano, and Auburger]{Nature17}
Lindsay~A Becker, Brenda Huang, Gregor Bieri, Rosanna Ma, David~A Knowles,
  Paymaan Jafar-Nejad, James Messing, Hong~Joo Kim, Armand Soriano, and Georg
  Auburger.
\newblock Therapeutic reduction of ataxin-2 extends lifespan and reduces
  pathology in tdp-43 mice.
\newblock \emph{Nature}, 544\penalty0 (7650):\penalty0 367--371, 2017.

\bibitem[Bl{\"o}schl et~al.(2019)Bl{\"o}schl, Hall, Viglione, Perdig{\~a}o,
  Parajka, Merz, Lun, Arheimer, Aronica, and Bilibashi]{Nature19}
G{\"u}nter Bl{\"o}schl, Julia Hall, Alberto Viglione, Rui~AP Perdig{\~a}o,
  Juraj Parajka, Bruno Merz, David Lun, Berit Arheimer, Giuseppe~T Aronica, and
  Ardian Bilibashi.
\newblock Changing climate both increases and decreases european river floods.
\newblock \emph{Nature}, 573\penalty0 (7772):\penalty0 108--111, 2019.

\bibitem[Chen et~al.(2001)Chen, Chen, and Kalbfleisch]{CheCK01}
Hanfeng Chen, Jiahua Chen, and John~D Kalbfleisch.
\newblock A modified likelihood ratio test for homogeneity in finite mixture
  models.
\newblock \emph{Journal of the Royal Statistical Society: Series B},
  63\penalty0 (1):\penalty0 19--29, 2001.

\bibitem[Chen et~al.(2004)Chen, Chen, and Kalbfleisch]{CheCK04}
Hanfeng Chen, Jiahua Chen, and John~D Kalbfleisch.
\newblock Testing for a finite mixture model with two components.
\newblock \emph{Journal of the Royal Statistical Society: Series B},
  66\penalty0 (1):\penalty0 95--115, 2004.

\bibitem[Dacunha-Castelle and Gassiat(1997)]{DacG97}
Didier Dacunha-Castelle and Elisabeth Gassiat.
\newblock Testing in locally conic models, and application to mixture models.
\newblock \emph{ESAIM: Probability and Statistics}, 1:\penalty0 285--317, 1997.

\bibitem[Dacunha-Castelle and Gassiat(1999)]{DacG99}
Didier Dacunha-Castelle and Elisabeth Gassiat.
\newblock Testing the order of a model using locally conic parametrization:
  population mixtures and stationary arma processes.
\newblock \emph{The Annals of Statistics}, 27\penalty0 (4):\penalty0
  1178--1209, 1999.

\bibitem[Daniel et~al.(2013)Daniel, Cousens, De~Stavola, Kenward, and
  Sterne]{DanCDKS13}
Rhian~M Daniel, S~N Cousens, B~L De~Stavola, Michael~G Kenward, and J~A~C
  Sterne.
\newblock Methods for dealing with time-dependent confounding.
\newblock \emph{Statistics in Medicine}, 32\penalty0 (9):\penalty0 1584--1618,
  2013.

\bibitem[Drton(2009)]{Drt09}
Mathias Drton.
\newblock Likelihood ratio tests and singularities.
\newblock \emph{The Annals of Statistics}, 37\penalty0 (2):\penalty0 979--1012,
  2009.

\bibitem[Efron et~al.(2004)Efron, Hastie, Johnstone, and Tibshirani]{EfrHJT04}
Bradley Efron, Trevor Hastie, Iain Johnstone, and Robert Tibshirani.
\newblock Least angle regression.
\newblock \emph{The Annals of Statistics}, 32\penalty0 (2):\penalty0 407--499,
  2004.

\bibitem[Fan and Li(2001)]{FanL01}
Jianqing Fan and Runze Li.
\newblock Variable selection via nonconcave penalized likelihood and its oracle
  properties.
\newblock \emph{Journal of the American Statistical Association}, 96\penalty0
  (456):\penalty0 1348--1360, 2001.

\bibitem[Fukumizu(2003)]{Fuk03}
Kenji Fukumizu.
\newblock Likelihood ratio of unidentifiable models and multilayer neural
  networks.
\newblock \emph{The Annals of Statistics}, 31\penalty0 (3):\penalty0 833--851,
  2003.

\bibitem[Gelfand et~al.(1992)Gelfand, Dey, and Chang]{GelDC92}
Alan~E Gelfand, Dipak~K Dey, and Hong Chang.
\newblock Model determination using predictive distributions with
  implementation via sampling-based method.
\newblock In \emph{Bayesian Statistics, 4th}, pages 147--167. Oxford University
  Press, 1992.

\bibitem[Gelman et~al.(2013)Gelman, Carlin, Stern, Dunson, Vehtari, and
  Rubin]{GelCSDVR13}
Andrew Gelman, John~B Carlin, Hal~S Stern, David~B Dunson, Aki Vehtari, and
  Donald~B Rubin.
\newblock \emph{Bayesian Data Analysis, 3rd}.
\newblock Chapman and Hall/CRC, 2013.

\bibitem[Imai and Ratkovic(2014)]{ImaR14}
Kosuke Imai and Marc Ratkovic.
\newblock Covariate balancing propensity score.
\newblock \emph{Journal of the Royal Statistical Society: Series B},
  76:\penalty0 243--263, 2014.

\bibitem[Knight and Fu(2000)]{KniF00}
Keith Knight and Wenjiang Fu.
\newblock Asymptotics for lasso-type estimators.
\newblock \emph{Annals of Statistics}, pages 1356--1378, 2000.

\bibitem[Konishi and Kitagawa(1996)]{KonK96}
Sadanori Konishi and Genshiro Kitagawa.
\newblock Generalised information criteria in model selection.
\newblock \emph{Biometrika}, 83\penalty0 (4):\penalty0 875--890, 1996.

\bibitem[Konishi and Kitagawa(2008)]{KonK08}
Sadanori Konishi and Genshiro Kitagawa.
\newblock \emph{Information Criteria and Statistical Modeling}.
\newblock Springer Science \& Business Media, 2008.

\bibitem[Korner-Nievergelt et~al.(2015)Korner-Nievergelt, Roth, Von~Felten,
  Gu{\'e}lat, Almasi, and Korner-Nievergelt]{KornerRVGAK15}
Franzi Korner-Nievergelt, Tobias Roth, Stefanie Von~Felten, J{\'e}r{\^o}me
  Gu{\'e}lat, Bettina Almasi, and Pius Korner-Nievergelt.
\newblock \emph{Bayesian Data Analysis in Ecology Using Linear Models with R,
  BUGS, and Stan}.
\newblock Academic Press, 2015.

\bibitem[Lv and Liu(2014)]{LvL14}
Jinchi Lv and Jun~S Liu.
\newblock Model selection principles in misspecified models.
\newblock \emph{Journal of the Royal Statistical Society: Series B}, pages
  141--167, 2014.

\bibitem[Matsuda et~al.(2021)Matsuda, Uehara, and Hyv{\"a}rinen]{MatUH21}
Takeru Matsuda, Masatoshi Uehara, and Aapo Hyv{\"a}rinen.
\newblock Information criteria for non-normalized models.
\newblock \emph{Journal of Machine Learning Research}, 2021.

\bibitem[Ninomiya and Kawano(2016)]{NinK16}
Yoshiyuki Ninomiya and Shuichi Kawano.
\newblock Aic for the lasso in generalized linear models.
\newblock \emph{Electronic Journal of Statistics}, 10\penalty0 (2):\penalty0
  2537--2560, 2016.

\bibitem[Robins et~al.(1994)Robins, Rotnitzky, and Zhao]{RobRZ94}
James~M Robins, Andrea Rotnitzky, and Lue~Ping Zhao.
\newblock Estimation of regression coefficients when some regressors are not
  always observed.
\newblock \emph{Journal of the American Statistical Association}, 89\penalty0
  (427):\penalty0 846--866, 1994.

\bibitem[Shibata(1989)]{Shi89}
Ritei Shibata.
\newblock Statistical aspects of model selection.
\newblock In \emph{From Data to Model}, pages 215--240. Springer, 1989.

\bibitem[Spiegelhalter et~al.(2002)Spiegelhalter, Best, Carlin, and Van
  Der~Linde]{SpiBCL02}
David~J Spiegelhalter, Nicola~G Best, Bradley~P Carlin, and Angelika Van
  Der~Linde.
\newblock Bayesian measures of model complexity and fit.
\newblock \emph{Journal of the Royal Statistical Society: Series B},
  64\penalty0 (4):\penalty0 583--639, 2002.

\bibitem[Stuart-Smith et~al.(2018)Stuart-Smith, Brown, Ceccarelli, and
  Edgar]{Nature18}
Rick~D Stuart-Smith, Christopher~J Brown, Daniela~M Ceccarelli, and Graham~J
  Edgar.
\newblock Ecosystem restructuring along the great barrier reef following mass
  coral bleaching.
\newblock \emph{Nature}, 560\penalty0 (7716):\penalty0 92--96, 2018.

\bibitem[Tibshirani(1996)]{Tib96}
Robert Tibshirani.
\newblock Regression shrinkage and selection via the lasso.
\newblock \emph{Journal of the Royal Statistical Society: Series B},
  58\penalty0 (1):\penalty0 267--288, 1996.

\bibitem[Tierney and Kadane(1986)]{TieK86}
Luke Tierney and Joseph~B Kadane.
\newblock Accurate approximations for posterior moments and marginal densities.
\newblock \emph{Journal of the American Statistical Association}, 81\penalty0
  (393):\penalty0 82--86, 1986.

\bibitem[Vehtari et~al.(2016)Vehtari, Mononen, Tolvanen, Sivula, and
  Winther]{VehMTSW16}
Aki Vehtari, Tommi Mononen, Ville Tolvanen, Tuomas Sivula, and Ole Winther.
\newblock Bayesian leave-one-out cross-validation approximations for gaussian
  latent variable models.
\newblock \emph{The Journal of Machine Learning Research}, 17\penalty0
  (1):\penalty0 3581--3618, 2016.

\bibitem[Vehtari et~al.(2017)Vehtari, Gelman, and Gabry]{VehGG17}
Aki Vehtari, Andrew Gelman, and Jonah Gabry.
\newblock Practical bayesian model evaluation using leave-one-out
  cross-validation and waic.
\newblock \emph{Statistics and Computing}, 27\penalty0 (5):\penalty0
  1413--1432, 2017.

\bibitem[Watanabe(2010)]{Wat10}
Sumio Watanabe.
\newblock Asymptotic equivalence of bayes cross validation and widely
  applicable information criterion in singular learning theory.
\newblock \emph{Journal of Machine Learning Research}, 11\penalty0 (12), 2010.

\bibitem[Watanabe(2013)]{Wat13}
Sumio Watanabe.
\newblock A widely applicable bayesian information criterion.
\newblock \emph{Journal of Machine Learning Research}, 14\penalty0
  (27):\penalty0 867--897, 2013.

\bibitem[Watanabe(2015)]{Wat15}
Sumio Watanabe.
\newblock Bayesian cross validation and waic for predictive prior design in
  regular asymptotic theory.
\newblock \emph{arXiv preprint: 1503.07970}, 2015.

\bibitem[Yuan and Lin(2006)]{YuaL06}
Ming Yuan and Yi~Lin.
\newblock Model selection and estimation in regression with grouped variables.
\newblock \emph{Journal of the Royal Statistical Society: Series B},
  68\penalty0 (1):\penalty0 49--67, 2006.

\bibitem[Zhang(2010)]{Zha10}
Cun-Hui Zhang.
\newblock Nearly unbiased variable selection under minimax concave penalty.
\newblock \emph{The Annals of Statistics}, 38\penalty0 (2):\penalty0 894--942,
  2010.

\end{thebibliography}

\end{document}